\documentclass[pra,twocolumn,showpacs,floatfix]{revtex4}
\usepackage{graphicx,amsmath,amsfonts,amssymb}
\usepackage[usenames]{color}

\begin{document}

\title{Localization of a spin-orbit coupled Bose-Einstein condensate in a bichromatic optical lattice}

\author{Yongshan Cheng$^{1}$\footnote{yong\_shan@163.com}}

\author{Gaohui Tang$^{1}$ \footnote{tgh19900201@163.com}}

\author{S. K. Adhikari$^2$\footnote{adhikari@ift.unesp.br;
URL: www.ift.unesp.br/users/adhikari}}

\affiliation{$^1$Department of Physics, Hubei Normal University,
Huangshi 435002,
People's   Republic of  China\\
$^2$Instituto de F\'{\i}sica Te\'orica, UNESP - Universidade
Estadual Paulista,
01.140-070 S\~ao Paulo, S\~ao Paulo, Brazil}

\date{\today}

\begin{abstract}

We study the localization of a  noninteracting and weakly interacting   Bose-Einstein condensate (BEC) with   spin-orbit coupling loaded in a quasiperiodic bichromatic optical lattice potential using the  numerical solution and variational approximation  of a binary mean-field  Gross-Pitaevskii equation with two  pseudo-spin components.
We confirm the existence of the stationary localized states   in the presence of the spin-orbit   and Rabi couplings
for an equal distribution of atoms in the two components. We find that the interaction between the spin-orbit  and Rabi couplings favors the localization or delocalization of the BEC depending on the the phase difference between the components. 
We also studied the oscillation dynamics of the localized states 
for an initial population imbalance between the two components.

\end{abstract}

\pacs{03.75.Mn, 03.75.Kk, 71.70.Ej, 72.15.Rn}

\maketitle

\section{Introduction}

The first experimental observation of spin-orbit (SO) coupling in Bose-Einstein condensates (BECs) \cite{nature-471-83} has stimulated widespread experimental and theoretical discussions in different fields. Spin-orbit coupled cold atoms represent a fascinating and fast developing area of research and  lead to rich physical effects \cite{nature-494-49}. In ultracold atomic systems, a variety of synthetic SO coupling can be engineered by two counter-propagating Raman lasers that couple two hyperfine ground states, and most experimental parameters can be controlled at will by optical or magnetic means \cite{experimental-scheme}. Using this technique, the SO coupling has been created in the atomic Fermi \cite{Fermi-gas} and Bose gases \cite{nature-471-83,Bose-gas}. Motivated by these experimental breakthroughs, a great number of theoretical activities have been devoted to the SO-coupled BEC including  superfluidity \cite{superfluidity}, vortex structure \cite{vortex}, and soliton \cite{soliton} of a SO-coupled BEC. There have also been extensive theoretical efforts toward understanding the physics of the SO-coupled Fermi gases \cite{BCS-BEC}. A generic binary mean-field Gross-Pitaevskii (GP) equation is derived in Refs. \cite{GP-equation1} which provide the starting point for the theoretical study of many-body dynamics in SO-coupled BECs. Similar models are also derived in other studies \cite{GP-equation2}, and have been employed in investigating the localized-modes \cite{localized-modes} and other topics of SO-coupled BECs \cite{use-GPE,use-GPE2}.

Another topic of current interest is the localization of a BEC in a disorder potential. Since Anderson predicted the localization of a noninteracting electron wave in solids with a disorder potential about $50$ years ago \cite{anderson}, localization phenomena have been studied in different types of waves including the atomic matter waves. Two experimental groups reported the localization of a noninteracting BEC in two different kinds of  one-dimensional (1D)  disordered potentials. Billy {\it et al.} \cite{billy} observed the exponential tail of the spatial density distribution of a $^{87}$Rb BEC after releasing it into a 1D waveguide with a controlled disorder potential created by a laser speckle. Roati {\it et al}. \cite{roati} observed the localization of a noninteracting $^{39}$K BEC in a 1D quasiperiodic bichromatic optical lattice (OL) potential. A  bichromatic OL is realized by a primary lattice perturbed by a weak secondary lattice with incommensurate wavelength \cite{PRL-98-130404}. The experimental realizations of three-dimensional (3D) localization of a spin-polarized Fermi gas  of $^{40}$K \cite{AL-3D1} atoms and a BEC of 
$^{87}$Rb \cite{AL-3D2} atoms in a 3D speckle potential were also reported. Much theoretical work has been done about Anderson localization of a BEC \cite{PRA-83-023620}. Recently, some theoretical investigations have been reported  for the localization  of a SO-coupled particle moving in a 1D quasiperiodic potential \cite{AL-in-SOC1} and random potential \cite{AL-in-SOC2}.

In this paper,  we investigate the statics and dynamics of  localization  of a noninteracting and weakly interacting   
BEC with Rabi and SO couplings,  trapped in a 1D quasiperiodic bichromatic OL potential $-$ similar to the one used in the experiment of Roati {\it et al.} \cite{roati} $-$ using a mean-field GP equation with two pseudo-spin components. The localized states are stationary for an equal occupation in the pseudo-spin states. But  when there is a population imbalance, spontaneous 
oscillation between the two pseudo-spin components of the localized state takes place. 
 We restrict ourselves to a study of the effect of the phase difference between the components, and of the SO and Rabi couplings on the statics and dynamics of 
localization. Within a range of parameters, most of the atoms can be  localized in a single  OL site and  the density profiles of the localized BEC are quite similar to a Gaussian shape, and the variational approximation can be used for some analytical understanding of the localized states \cite{variational}. In view of the SO coupling, for the initial ansatz of the wave function  we choose in our analysis has a somewhat more complicated form in order to get an understanding of the characteristic of the localized BECs.   The stability criterion of the stationary localized states is discussed by performing a standard linear stability analysis. We also study the tails of the localized states, where we focus on the spatially extended nature of  wave functions with exponential decay corresponding to a weak Anderson localization \cite{anderson,review-AL}.  We also study  the dynamics of atom transfer between the two localized components with a population imbalance.

In Sec. \ref{II} we present a brief account of the coupled mean-field model and the bichromatic OL potential used in the study. The analytical expressions for the atom transfer ratio and phase difference between the two localized states, and width of the two localized states are obtained by the variational analysis of the mean-field model. Various aspects of stationary localization are studied by variational approximation and numerical solution of the mean-field equation. In particular, the effect of the phase difference between components and of the SO and Rabi couplings on the localization of the BEC are investigated in Sec. \ref{III}. Some dynamics of the nonstationary localized states are presented  in Sec. \ref{IIII}. A brief summary and future perspective are given in Sec. \ref{IIIII}.

\section{Analytical consideration }

\label{II}

{In electronic states of an atom the SO coupling naturally appears due to the magnetic energy associated with this coupling because of the electronic charge. In the case of neutral atoms an engineering with electromagnetic fields is required for the SO coupling to contribute to the BEC.     
To create a simple  SO coupling in the laboratory, Lin {\it et al.}
\cite{nature-471-83}  consider two internal spin states of $^{87}$Rb  hyperfine state 
5S$_{1/2}$:   
$|\uparrow\rangle = |F=1, m_F=0\rangle $ and  $|\downarrow \rangle = |F=1, m_F=-1\rangle $ 
  where $F$ and $m_F$ are the total angular momentum of the hyperfine state and its $z$ projection.
These states are called pseudo-spin-up and pseudo-spin-down  states in analogy with the two 
spin components of a spin half particle. }
The SO coupling between these states 
is then realized with strength $\Omega$ using two counterpropagating Raman lasers 
and this SO coupling is equivalent to that of an electronic system with equal contribution of Rashba \cite{7} and Dresselhaus \cite{8} couplings and with an external uniform magnetic field.
We consider a BEC   with internal up and down pseudo-spin states  $|\rm{\uparrow}\rangle$ and  $|\rm{\downarrow}\rangle$ confined in a spin-independent quasi-1D potential $V(x)$ oriented in the longitudinal ($x$) direction. A strong  harmonic potential of angular frequency $\omega_\perp$ is applied in transverse $y, z$ directions, and the transverse dynamics of the condensate is assumed to be  frozen to the respective ground states of harmonic traps. 
Then, the single-particle quasi-1D Hamiltonian of the system under the action of a strong transverse trap of angular frequency $\omega_\perp$ in the $y-z$ plane 
can be written as \cite{nature-471-83,localized-modes,use-GPE}
\begin{eqnarray}\label{Hamilt}
H_0={ \frac{p_x^2}{2m}}+\frac{\hbar k_L}{m} p_x\sigma_z+{\frac{\hbar \Omega}{2}}\sigma_x+V(x),
\end{eqnarray}
where $p_x = - i\hbar \partial_x$ is the momentum operator along  $x$ direction, $m$ is the mass of an atom, $\sigma_{x,z}$ are the usual $2\times 2$ Pauli matrices,  $k_L$ is the wave number of the Raman lasers that  couple the two atomic hyperfine states, and the couplin strength $\Omega$ is the Rabi frequency  acting  as a Zeeman field. 
If the interactions among the atoms in the BEC are taken into account, in the Hartree approximation, the dynamics of the BEC of $N$ atoms  can be described by the $2\times 2$ nonlinear 1D GP equation \cite{soliton,GP-equation1,GP-equation2}:
\begin{eqnarray}\label{eq1d}
i\hbar\frac{\partial \psi}{\partial t}=H_0 \psi + G\psi,
\end{eqnarray}
where $\psi=(\psi_1, \psi_2)^T$ is the two-component mean-field wave function with normalization $\int dx (|\psi_1|^2+|\psi_2|^2) =N$. The two time-dependent spinor wave functions $\psi_j (j=1,2)$ describe  the two pesudo-spin components ($|\rm{\uparrow}\rangle$ and $|\rm{\downarrow}\rangle$) of the BEC. The nonlinear term has the $2\times 2$ matrix form \cite{GP-equation1,1dsala}
\begin{eqnarray}\label{scat}
G=\left( \begin{array}{cc} \frac{2\hbar^2a|\psi_1|^2+2\hbar^2a_{12}|\psi_2|^2}{ma_\perp^2} &
0\\ 0 & \frac{2\hbar^2a|\psi_2|^2+2\hbar^2a_{12}|\psi_1|^2|}{ma_\perp^2} \\
\end{array}\right),
\end{eqnarray}
where, to make the parameters of the model tractable, we take the two intraspecies scattering lengths $a_j$ to be equal: $a_1=a_2=a$, and where $a_{12}$ is the interspecies scattering length, and $a_{\perp}=\sqrt{\hbar/(m\omega_\perp)}$ is the   harmonic oscillator length of the transverse trap.  In actual experiment it is possible to control these scattering lengths independently by optical \cite{opt} and magnetic \cite{Feshbach} Feshbach resonance techniques. 
In dimensionless units the coupled GP equations for the wave function 
{ $u_j\equiv u_j(x,t)= \psi_j(x,t)\sqrt{a_\perp}$}  $(j=1,2)$ can be written as
\cite{localized-modes}:
\begin{eqnarray}\label{CGP1}
i\frac{\partial u_j}{\partial t}=- \frac{1}{2}\frac{\partial^2
u_j}{\partial x^2} +i(-1)^j\gamma \frac{\partial u_j}{\partial x}
+\Gamma u_{(3-j)}
\nonumber  \\
+ (g|u_j|^2+g_{12}|u_{(3-j)}|^2)u_j +
V(x)u_j,
\end{eqnarray}
where the spatial variable $x$, time $t$, density $|u_j|^2$, and energy are expressed in normalized  units $a_\perp$, $\omega_\perp^{-1}$, $a_\perp^{-1}$ and $\hbar\omega_\perp$, respectively.   The interaction nonlinearities are \cite{1dsala} $g=2a/a_\perp^2, g_{12}=2a_{12}/a_\perp^2$, the SO-coupling strength is  $\gamma\equiv k_L a_\perp$ and the Rabi-coupling strength is
{ $\Gamma\equiv\Omega/(2\omega_\perp)$}. The normalization $\int_{-\infty}^\infty  |u_j|^2dx=N_j$, where $N_j$ is the number of atoms in component $j$.  
As in the experiment of Roati {\it et al.} \cite{roati}, the  bichromatic OL potential $V(x)$ is taken as the linear combination of two polarized standing wave OL potentials of incommensurate wave lengths:
\begin{eqnarray}\label{pot}
V(x)=\sum_{l=1}^2 A_l \sin^2(k_lx),
\end{eqnarray}
with $A_l=2\pi^2 s_l/\lambda_l^2, (l=1,2)$, where $\lambda_l$'s are the wavelengths of the OL potentials, $s_l$'s are their intensities, and $k_l=2\pi/\lambda_l$ the corresponding wave numbers. In this investigation, the irrational ratio between the two OL is set to be  \cite{PRA-83-023620} $k_2/k_1=(\sqrt{5}-1)/2$, the inverse of the golden ratio. In the actual experiment of Roati {\it et al.} \cite{roati},  the parameter was set as: $k_2/k_1=1.1972$. Without losing generality, we further take $\lambda_1=10$, and $s_1=10$, $s_2=0.3s_1$ which are roughly the same parameters as in the experiment of Roati {\it et al.} \cite{roati}.

The dynamics of the BEC  can be investigated by the Gaussian variational approach \cite{variational}. { This approach  is justified for small contact repulsion  and for small  
 SO coupling, when the localized state has a spatial extention 
over a single OL site. In such a situation, the central density of the localized state has an 
approximate Gaussian shape. However, at large distances the localized state has a long 
exponential tail. The Gaussian variational approach  can describe the Anderson localization experiment of Ref. \cite{roati} in the 
noninteracting regime.}
 In this approach, the Lagrangian density for Eq. (\ref{CGP1})  is
\begin{align}\label{den1}
{\cal L}&=\sum_{j=1}^2\biggr\{\frac{i}{2}\left(u_j^*\dot{u}_j-u_j\dot{u}_j^*\right)
-(-1)^j\frac{i}{2}\gamma\left[u_j^*u_j'-u_j(u_j^*)'\right]
\nonumber\\
&-\frac{1}{2}|u_j'|^2-\frac{1}{2}g|u_j|^4-V(x)|u_j|^2\biggr\}
-g_{12}|u_1|^2|u_2|^2\nonumber\\
&-\Gamma\left(u_1^*u_2+u_1u_2^*\right),
\end{align}
where the star denotes the complex conjugate, the prime denotes $d/dx$, and the overhead dot denotes $d/dt$. The Gaussian  ansatz with the {\it time-dependent} variational parameters $N_j, w_j, \beta_j$ and $\phi_j$  is used to study  the dynamics:
\begin{eqnarray}\label{ans}
u_j(x,t)=\frac{1}{\pi^{1/4}}\sqrt{\frac{N_j}{w_j}}\exp\left[-\frac{ x^2} {2w_j^2}+ (-1)^ji\beta_j x+i\phi_j\right],
\end{eqnarray}
where $N_j, w_j$ represent the number of atoms and width of the BEC, and
  $\beta_j$ and $\phi_j$ are chirp and phase. { The time-dependence 
of these variables is not explicitly shown in the following.}
  The effective Lagrangian of the system (\ref{CGP1}) is found by substituting Eq. (\ref{ans}) into Eq. (\ref{den1}) and
integrating over space variables \cite{variational}:
\begin{widetext}
\begin{eqnarray}\label{Lag}
L&=&\sum_{j=1}^2N_j\biggr[-\dot{\phi}_j+\gamma\beta_j-\frac{1}{2}\left(\frac{1}{2w_j^2}+\beta_j^2
+\frac{gN_j}{\sqrt{2\pi}w_j}\right)
 +\frac{1}{2}\sum_{l=1}^2A_l\left[\exp(-k_l^2w_j^2)-1\right]\biggr]
-\frac{g_{12}N_1N_2}{\sqrt{\pi(w_1^2+w_2^2)}}\nonumber \\
&-&2\Gamma\cos(\phi_2-\phi_1)\sqrt{N_1N_2}L_\Gamma,
\end{eqnarray}
\end{widetext}
where
\begin{align}  \label{LGM}
L_\Gamma &= \sqrt{\frac{2w_1w_2}{w_1^2+w_2^2}}\exp\left[-\frac{(\beta_1+\beta_2)^2w_1^2w_2^2}{2(w_1^2+w_2^2)}\right].
\end{align}

We further define atom transfer ratio between the two localized states: $R=(N_2-N_1)/N$, and  the phase difference: $\phi\equiv\phi_2-\phi_1$. Using the Euler-Lagrange equation
\begin{eqnarray}\label{EUL}
\frac{\partial L}{\partial \alpha}-\frac{d}{dt}\frac{\partial L}{\partial \dot{\alpha}}=0,
\end{eqnarray}
where $\alpha$ denotes the variational parameters $\phi_j, N_j, w_j$ and $\beta_j$, respectively, we obtain the following equations
\begin{widetext}
\begin{eqnarray}
\label{PHI}
\dot{\phi}&=&\gamma\left(\beta_2-\beta_1\right)
-\frac{1}{4}\left(\frac{1}{w_2^2}-\frac{1}{w_1^2}\right)
-\frac{gN}{2\sqrt{2\pi}}\left(\frac{1+R}{w_2}-\frac{1-R}{w_1}\right)
-\frac{1}{2}\left(\beta_2^2-\beta_1^2\right)
\nonumber\\
&&+\frac{1}{2}\sum_{l=1}^2A_l\left[\exp\left(-k_l^2w_2^2\right)
-\exp\left(-k_l^2w_1^2\right)\right]+\frac{g_{12}RN}{\sqrt{\pi(w_1^2+w_2^2)}}
+\frac{2R\Gamma\cos\phi}{\sqrt{1-R^2}}L_\Gamma \equiv G(R,\phi),\\\label{WID}
0&=&\frac{1}{2w_j^3}+\frac{gN_j}{2\sqrt{2\pi}w_j^2}-w_j\sum_{l=1}^2A_lk_l^2\exp\left(-k_l^2w_j^2\right)+\frac{g_{12}N_{3-j}w_j}{\sqrt{\pi(w_1^2+w_2^2)^3}}-2\Gamma \cos \phi\sqrt{\frac{N_{3-j}}{N_j}}\frac{\partial L_\Gamma}{\partial w_j},
\end{eqnarray}
\end{widetext}
\begin{align}
\label{Nj}
\dot{R}&=-2\Gamma\sin \phi\sqrt{1-R^2}L_\Gamma\equiv F(R,\phi),\\
 \label{BET}
0&=\gamma-\beta_j+2\Gamma \cos \phi\sqrt{\frac{N_{3-j}}{N_j}}\frac{(\beta_2+\beta_1)w_1^2w_2^2}{w_1^2+w_2^2}L_\Gamma.
\end{align}
Equation (\ref{Nj}) shows that the transfer ratio $R$ is explicitly dependent on $\Gamma$, which implies that the Rabi coupling leads to atom transfer between the two  localized states.

\section{STATIONARY LOCALIZED STATE}

\label{III}
The stationary states are obtained by setting the time derivative in Eqs.  (\ref{PHI}) and  (\ref{Nj})  to zero. If $R=0$, i.e., $N_1=N_2$, we obtain $\beta_1=\beta_2\equiv\beta$ from Eq. (\ref{BET}),  $w_1=w_2\equiv w$ from Eq. (\ref{WID}), and  $\dot{\phi}=0$ from Eq. (\ref{PHI}).
Consequently,
the two localized states given by Eq. (\ref{ans}) are identical. Hence, the simple stationary solutions, in this case, are
{ \begin{eqnarray}\label{SOL1}
R&=&0, \quad \phi=0 \quad (\text{in phase})\\ \label{SOL2}
R&=&0, \quad \phi=\pi \quad (\text{out of phase}).
\end{eqnarray}
}
Hence, Eqs. (\ref{WID}) and (\ref{BET}) can be rewritten as
\begin{eqnarray}\label{WID1}
0&=&\frac{1}{2w^3}+N\frac{g+g_{12}}{4\sqrt{2\pi}w^2}-w\sum_{l=1}^2A_lk_l^2\exp(-k_l^2w^2)\nonumber\\
&&\pm2\Gamma w\beta^2\exp(-\beta^2w^2),\\ \label{BET1}
0&=&\gamma-\beta\pm2\Gamma w^2\beta\exp(-\beta^2w^2),
\end{eqnarray}
where ``$+$'' corresponds to $\phi=0$, and ``$-$'' corresponds to $\phi=\pi$. We find that the stationary states are related to the external trapping potential, nonlinearity, phase difference, SO and  Rabi couplings. 
In order to focus our attention on the effects of the phase difference and  SO and Rabi couplings on the localization of the BEC, here, we will restrict ourselves first  to the noninteracting regime. The weakly interacting regime, and even the noninteracting one, could be achieved by reducing the s-wave scattering length by means of Feshbach resonances \cite{Feshbach}. We take $g=g_{12}=0$ with potential   (\ref{pot}) in the following investigations.

 We solve Eq. (\ref{CGP1})  by the real- or imaginary-time split-step Fourier spectral method with a space step 0.04 and time step 0.001.  In real-time propagation, to obtain the stationary localized states,
we take  the stationary solution of Eq. (\ref{CGP1}) for $g=g_{12}=\Gamma=0$ and $V(x) = x^2/2$, e.g.  $u_j(x)=\pi^{-1/4}\exp[-x^2/2+(-1)^j i\gamma x+i\phi_{j0}]$, as the initial input. Successively, the parabolic trap is slowly turned off and the bichromatic OL is slowly turned on and the parameter $\Gamma$ is added gradually in steps of $0.000001$
 from $0$ to the final value.   To investigate the effects of the phase difference, we take $\phi_{10}=\phi_{20}=0$ for the in-phase case, $\phi_{10}=0$ and $\phi_{20}=\pi$ for the out-of-phase configuration 
in the initial input pulses.

\begin{figure} 
\begin{center}
\includegraphics[width=.49\linewidth]{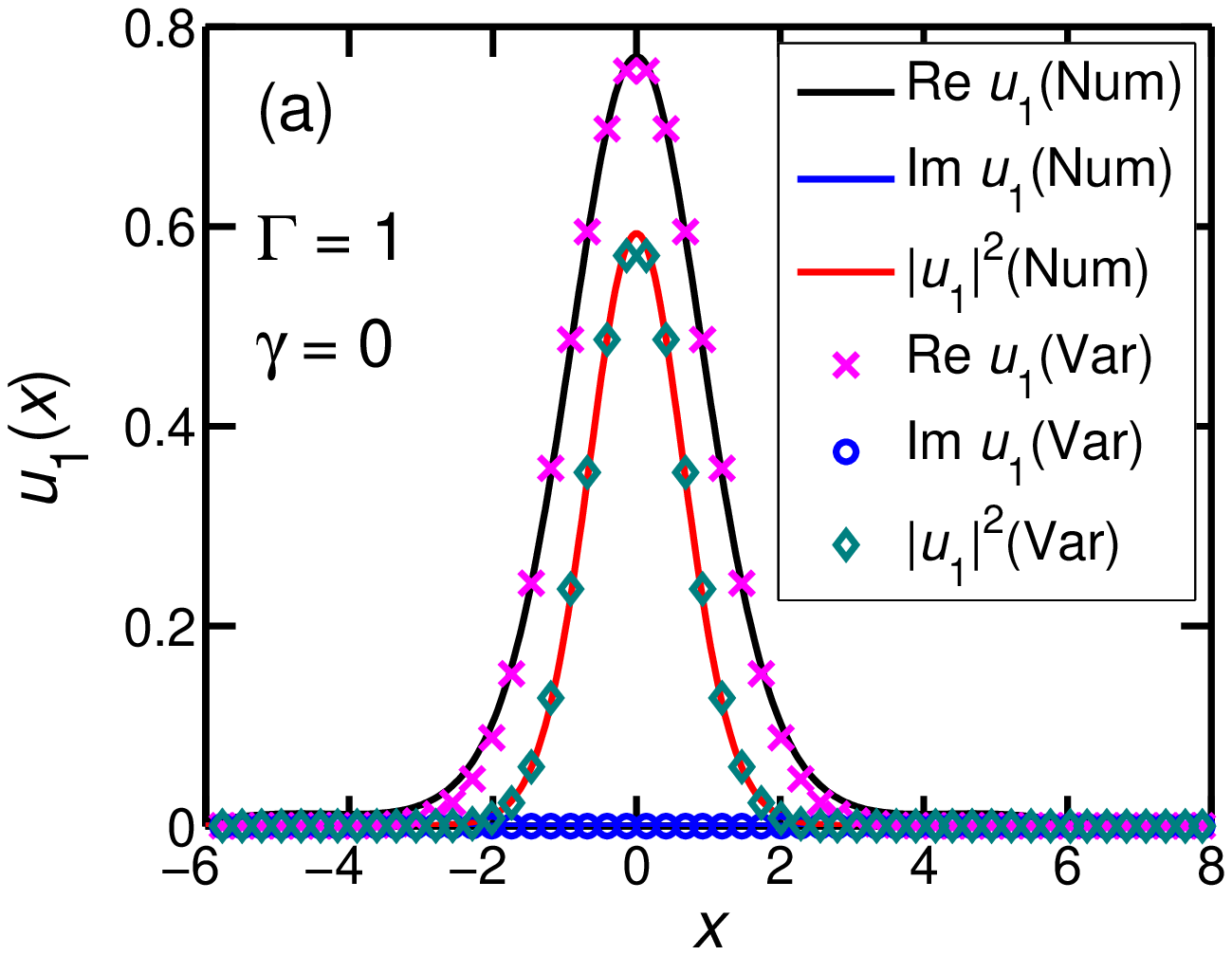}
\includegraphics[width=.49\linewidth]{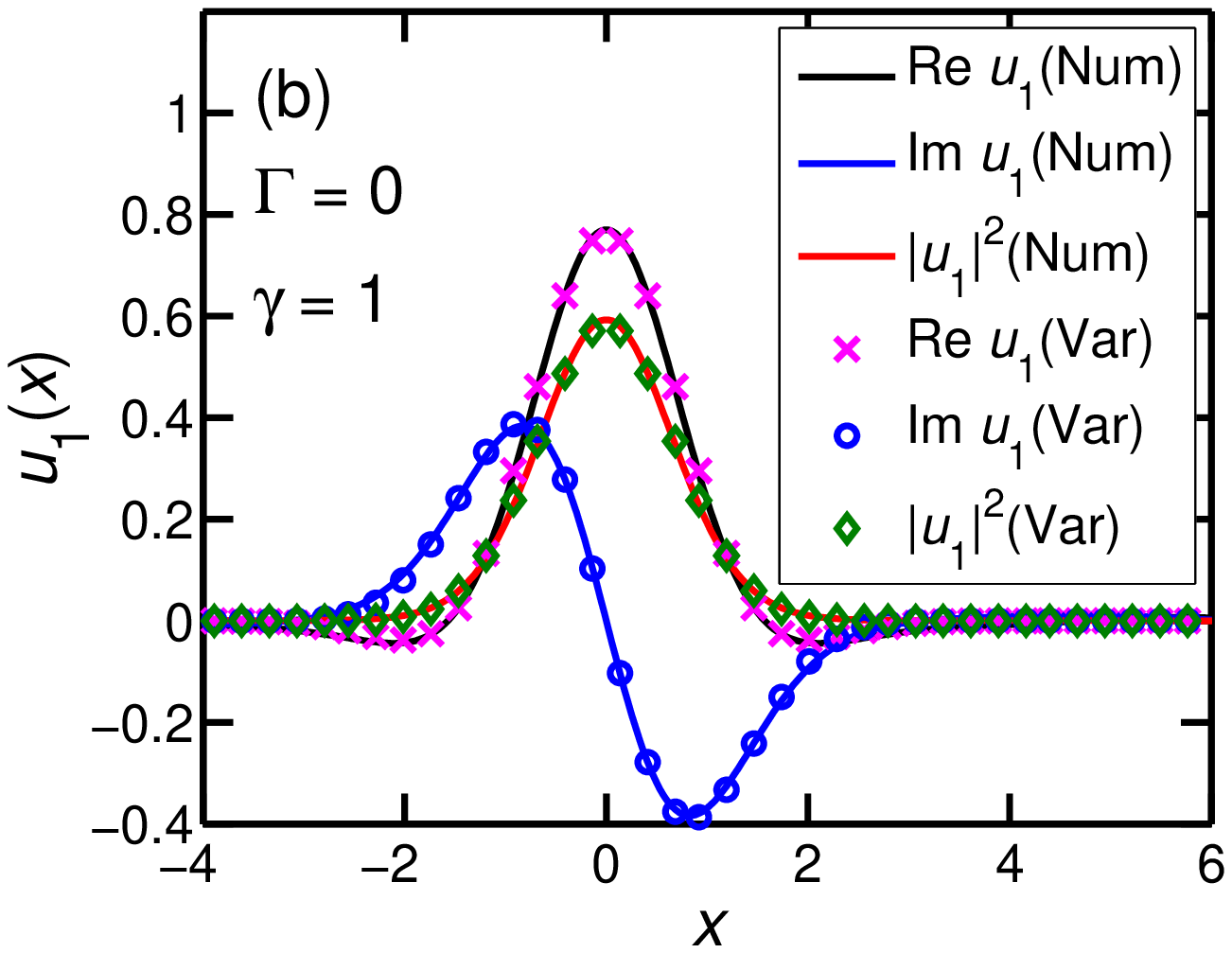}
\includegraphics[width=.49\linewidth]{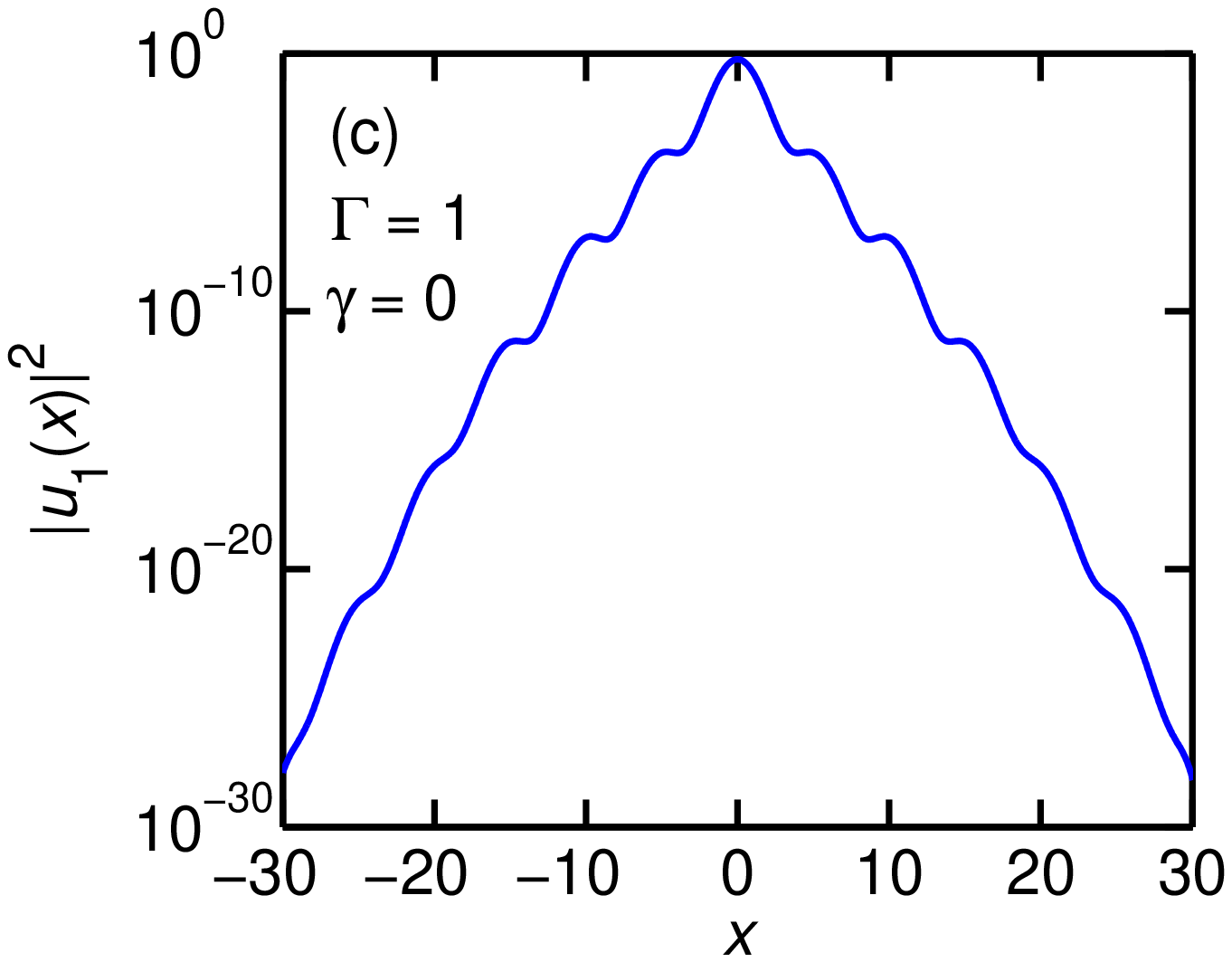}
\includegraphics[width=.49\linewidth]{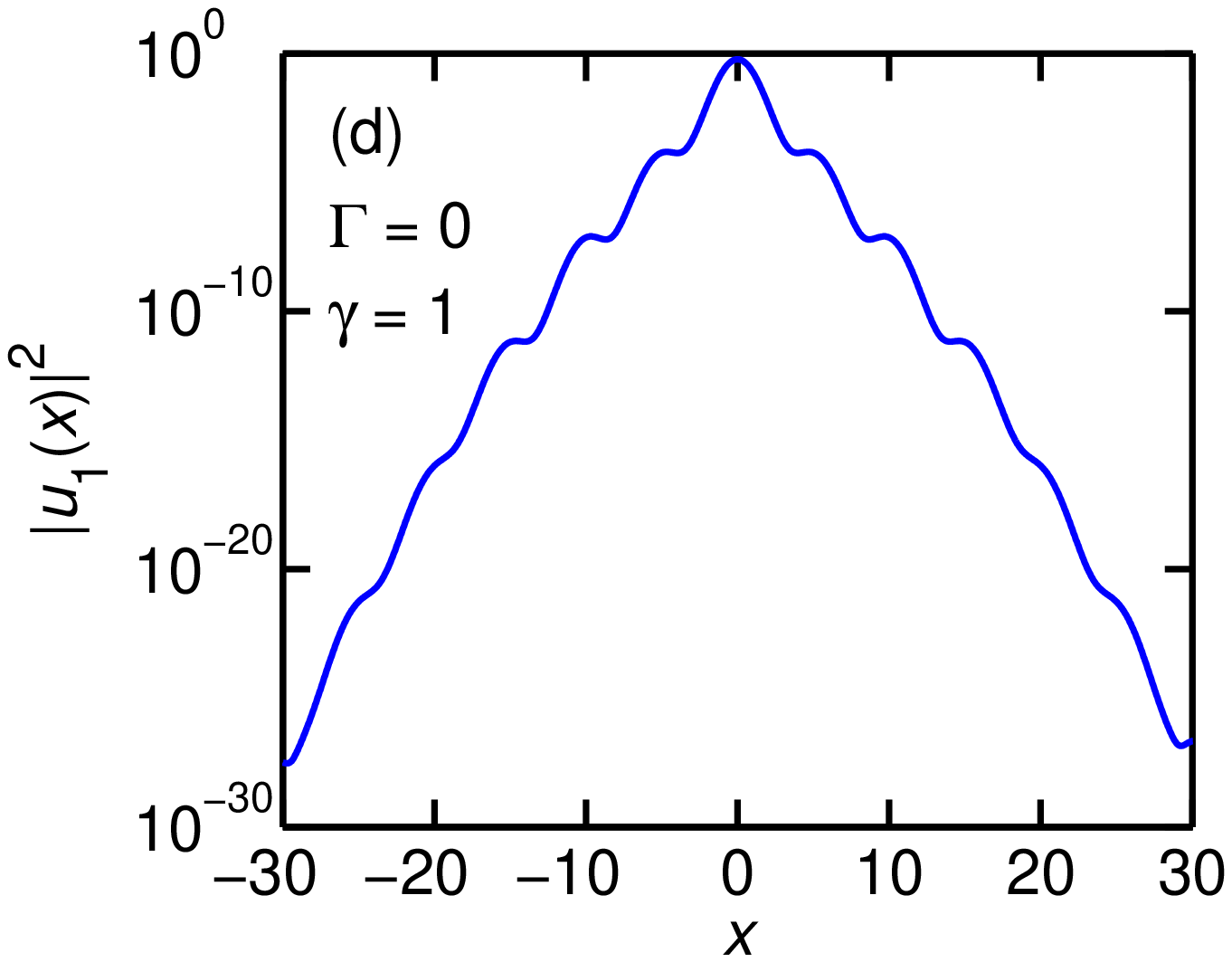}
\end{center}

\caption{(Color online) Real (Re) and imaginary (Im) parts of numerical (Num) and variational (Var) wave function $u_1(x)$ of Eq. (\ref{CGP1}) with the bichromatic OL potential (\ref{pot})
and the corresponding density $|u_1(x)|^2$
 versus $x$ for $g=g_{12}=0$ and
(a) $\Gamma =1, \gamma =0, N_1=N_2=1$ and (b) $\Gamma =0, \gamma =1$.
The corresponding { numerical} densities $|u_1(x)|^2$  on log scale are shown in (c) and (d).
Because of the similarity between $u_1$ and $u_2$, only the wave function $u_1$ is plotted. All quantities are dimensionless. (similarly here after).
 } \label{fig1}
\end{figure}

In this section for the calculation of stationary states we take throughout $N_1=N_2=1.$
Let us, first, investigate the effects of the coefficient $\gamma$ and $\Gamma$ on the noninteracting localized states when $\gamma\times\Gamma=0$ and $g=g_{12}=0$.  If $\gamma=0$, a  solution of Eq. (\ref{BET1}) is $\beta=0$. Then, the last term on the right-hand side of Eq. (\ref{WID1}) is zero, so that the widths are not related to the phase difference and Rabi coupling $\Gamma$. Also, if $\Gamma =0$, Eq. (\ref{WID1}) shows that the widths are independent of the phase difference $\phi$ and SO coupling $\gamma$. In both cases the widths are determined by only the external trapping potential and nonlinearity. In these cases, the variational width is $0.9688$, and the numerical width ($w^2=2\int_{-\infty}^{+\infty}x^2|u|^2dx$) is $0.9945$. The numerical simulation of Eq. (\ref{CGP1}) shows that $u_1$ and $u_2$ are similar: $|u_1|^2=|u_2|^2$, but $\text{Im}(u_1)=-\text{Im}(u_2)$ and $\text{Re}(u_1)=\text{Re}(u_2)$ for $\phi_{10}=\phi_{20}=0$, and  $\text{Im}(u_1)=\text{Im}(u_2)$ and $\text{Re}(u_1)=-\text{Re}(u_2)$ for $\phi_{10}=0$ and $\phi_{20}=\pi$. Because of the similarity between $u_1$ and $u_2$, we plot only $u_1$ here after,  as  in Figs. \ref{fig1} (a) and (b),  illustrating the variational and numerical results for the localized states.   The variational wave function is obtained   by solving Eqs. (\ref{WID1})  and  substituting $w$ and $\beta$ into Eq. (\ref{ans}). We find that the variational results are in good agreement with the numerical results.

Anderson localization in a weakly disordered potential is characterized by a long exponential tail of the localized state \cite{anderson,review-AL}. To observe the effects of the SO and Rabi couplings on the tail region, we plot in Figs. \ref{fig1} (c) and \ref{fig1} (d) the density distribution $|u_1|^2$ of the stationary BEC on log scale.  The parameters   $\gamma$ and $\Gamma$ have no effect on the exponential tail when $\gamma\times\Gamma=0$, as confirmed by the numerical simulation of Eq. (\ref{CGP1}) with different  $\gamma$ and $\Gamma$.

Next, we consider $\gamma\times\Gamma\neq 0$. Because of the interaction between $\gamma$ and $\Gamma$, now the width of the stationary state should depend on the the phase difference, SO and Rabi couplings.   However,
Eqs. (\ref{WID1}) and (\ref{BET1}) show that
the width of the  localized state is determined by the difference $(\gamma-\beta)$.

\begin{figure}
\begin{center}
\includegraphics[width=.49\linewidth]{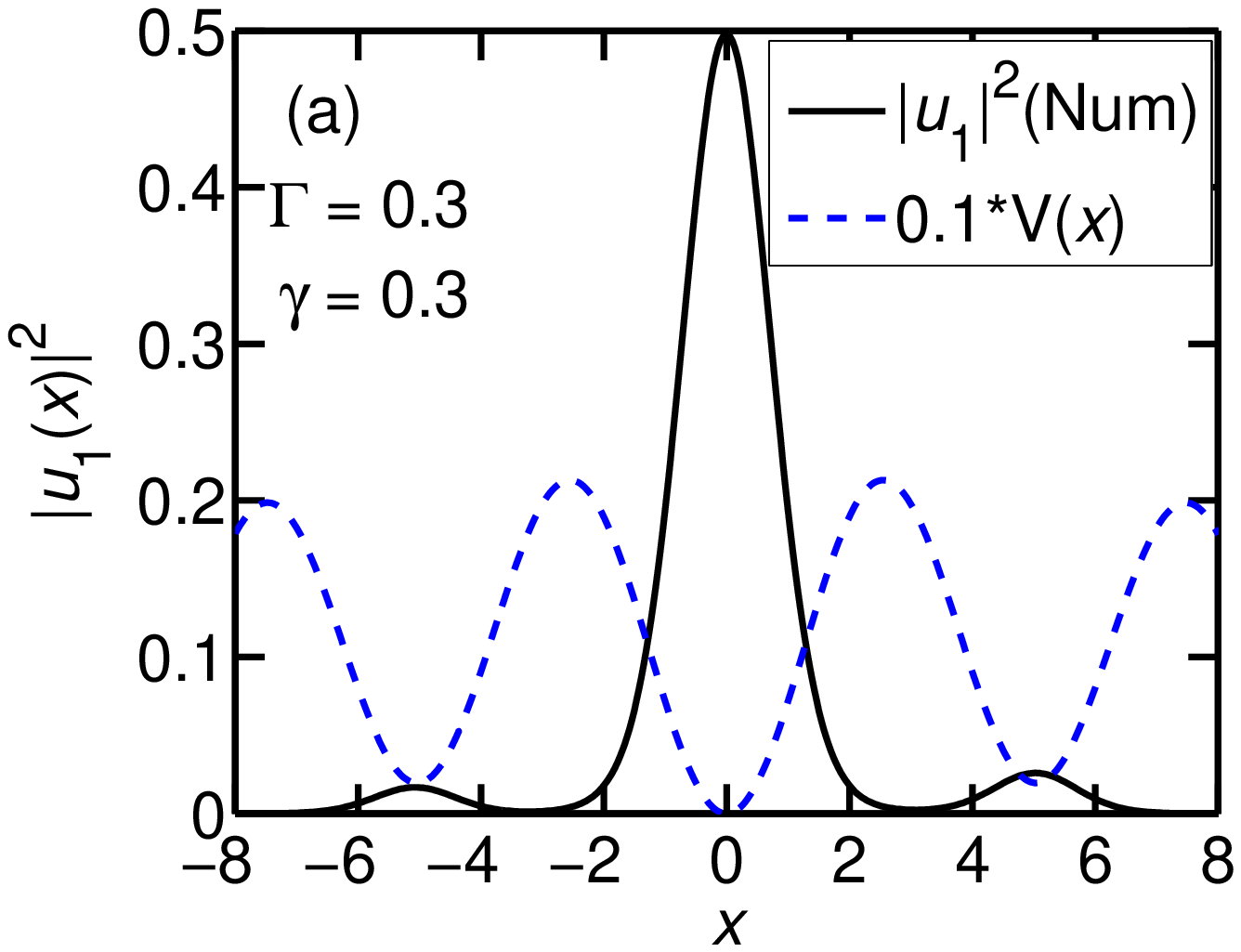}
\includegraphics[width=.49\linewidth]{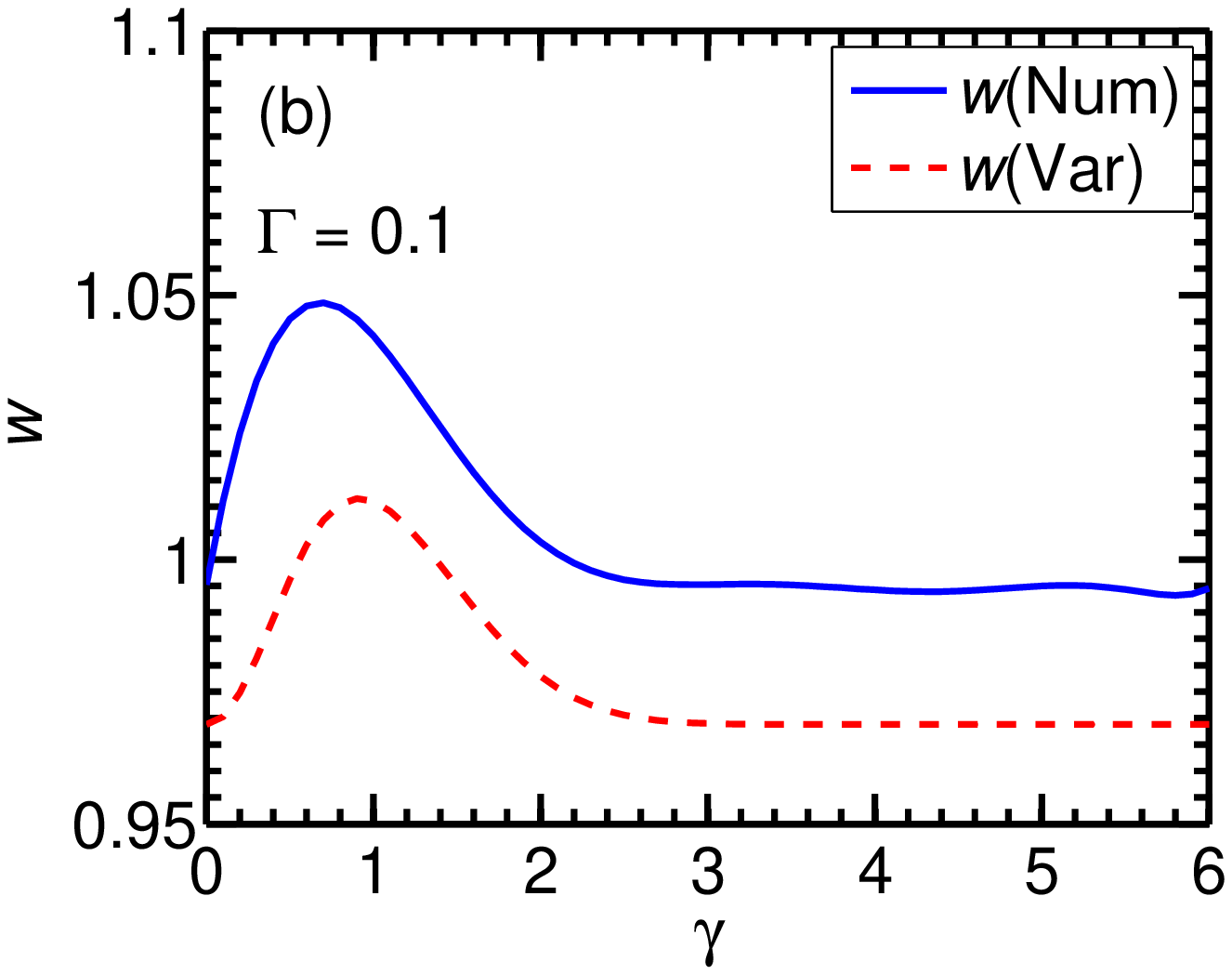}
\end{center}

\caption{(Color online)  Numerical and variational results for $\phi=0$ (in phase), $g=g_{12}=0, N_1=N_2=1$  (a) of density  $|u_1|^2$ versus $x$ for $\gamma=\Gamma=0.3$ and  (b) of   width versus $\gamma$ for $\Gamma= 0.1$. To compare, the OL potential $0.1V(x)$ is also plotted in (a).
 } \label{fig2}
\end{figure}

{ In the case of $\phi=0$ (in phase), the last term in 
Eq. (\ref{WID1}) is positive for 
 positive Rabi coupling $\Gamma$ and  contributes to a delocalization of the BEC as the positive 
kinetic energy term (the first term on the right hand side) and the positive repulsive  
interaction term (the second term on the right hand side). The only  term contributing 
to localization is the negative bichromatic OL term (the third term on the right hand 
side) in Eq. (\ref{WID1}).
Hence a large positive $\Gamma$   should lead to a partially delocalized state 
occupying a large spatial region extending over several OL sites. Such a localized state 
over multiple OL sites has  
a multi-hump structure.} To acquire a single-hump localized state, Rabi coupling $\Gamma$ must be small.
 Then, it follows  from Eq. (\ref{BET1}) that if
  $\gamma$  is large enough,  $\beta\simeq\gamma$ and the width is independent of $\beta, \gamma$ and $\Gamma$. For example, we obtain $\beta=3.5, \; w=0.9688$ by numerically solving Eqs. (\ref{WID1}) and (\ref{BET1}) with $\gamma=3.5,\;\Gamma=0.1$.  The numerical integration of Eqs. (\ref{CGP1}) and (\ref{pot}) shows that a single-humped localized state splits into a multi-humped state occupying more than one OL site with the increase of
 $\Gamma$, as illustrated in  Fig. \ref{fig2} (a) for  $\gamma=\Gamma= 0.3$.   If $\Gamma=0.1$ is small, the density profile occupies only one OL site within a wide range of parameter $\gamma$.   In Fig. \ref{fig2} (b) we compare the numerical and variational widths of the localized state for $\Gamma=0.1$ and different $\gamma$ and find that the width is the smallest for $\gamma=0$   implying that  a positive $\gamma$ contributes to a slight delocalization for small $\gamma$.  For larger $\gamma$  ($>3$), the width is practically independent of $\gamma$. These findings are consistent with  variational Eqs. (\ref{WID1}) and (\ref{BET1}). In Fig. \ref{fig2} (b) the numerical width is slightly larger than the variational width consistent with  a long exponential tail of the former.   We  also studied the tails of the density profiles in logarithmic scale, and found that the effect of $\gamma$ on the tails is small.

\begin{figure}
\begin{center}
\includegraphics[width=.49\linewidth,clip]{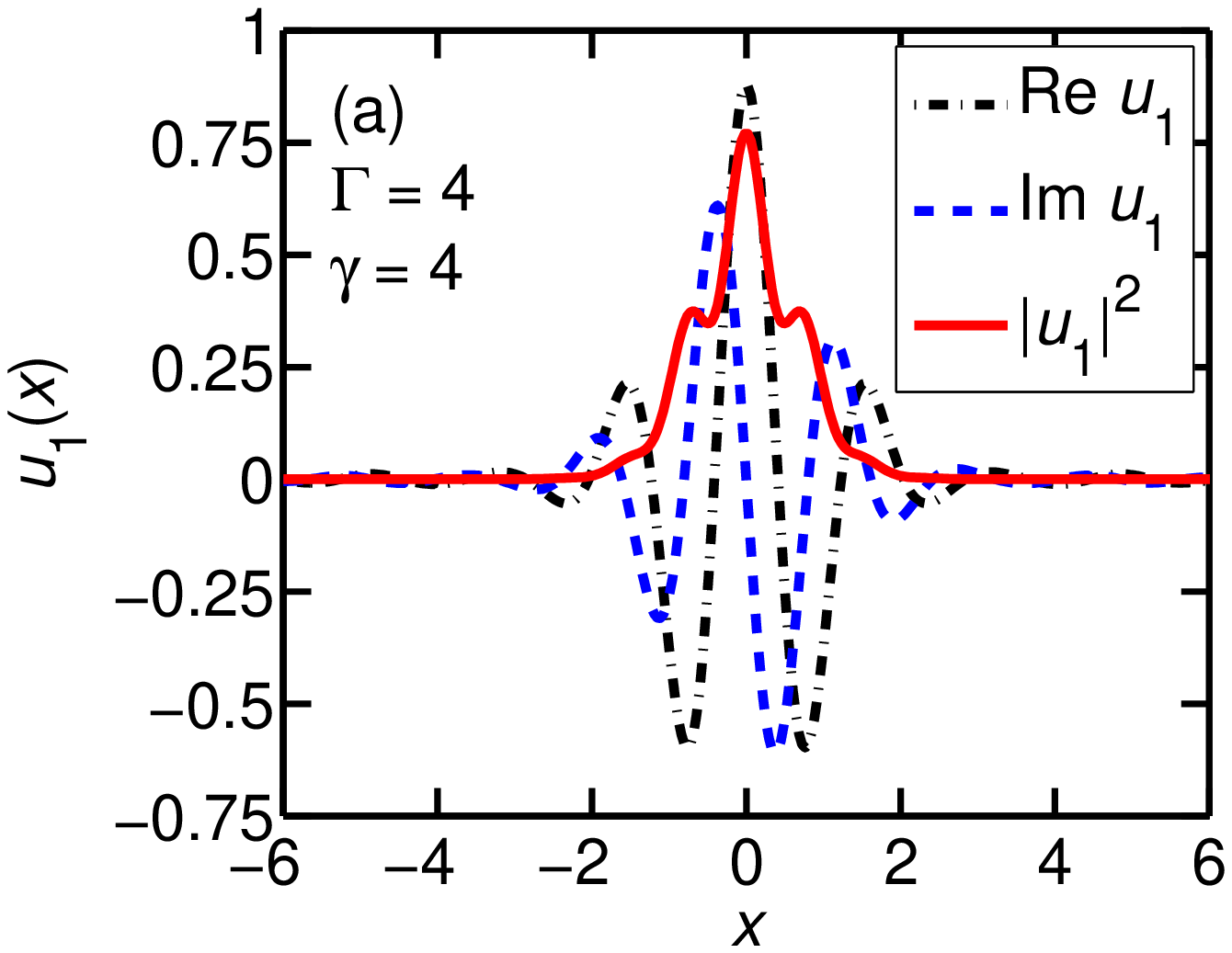}
\includegraphics[width=.49\linewidth,clip]{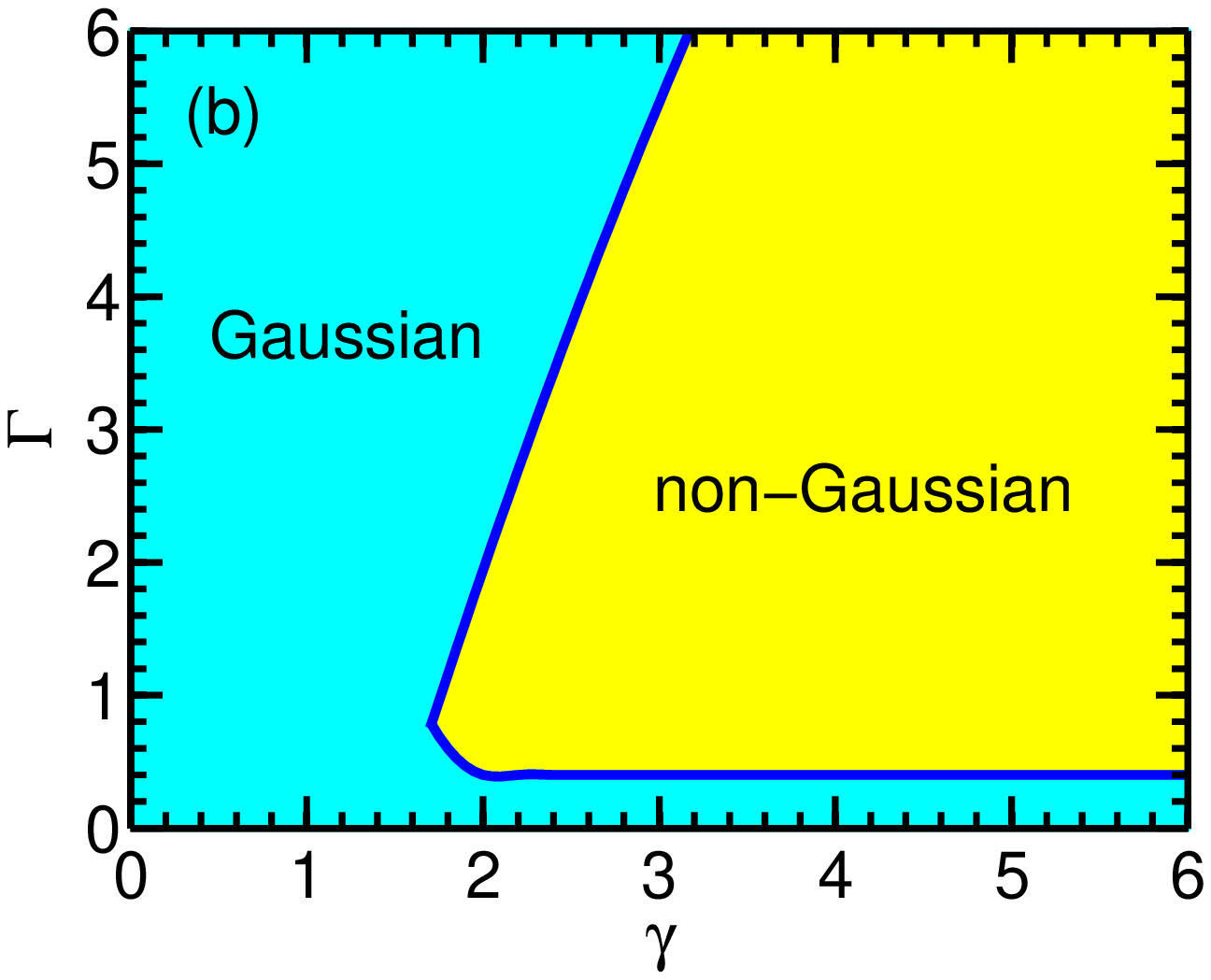}
\includegraphics[width=.49\linewidth,clip]{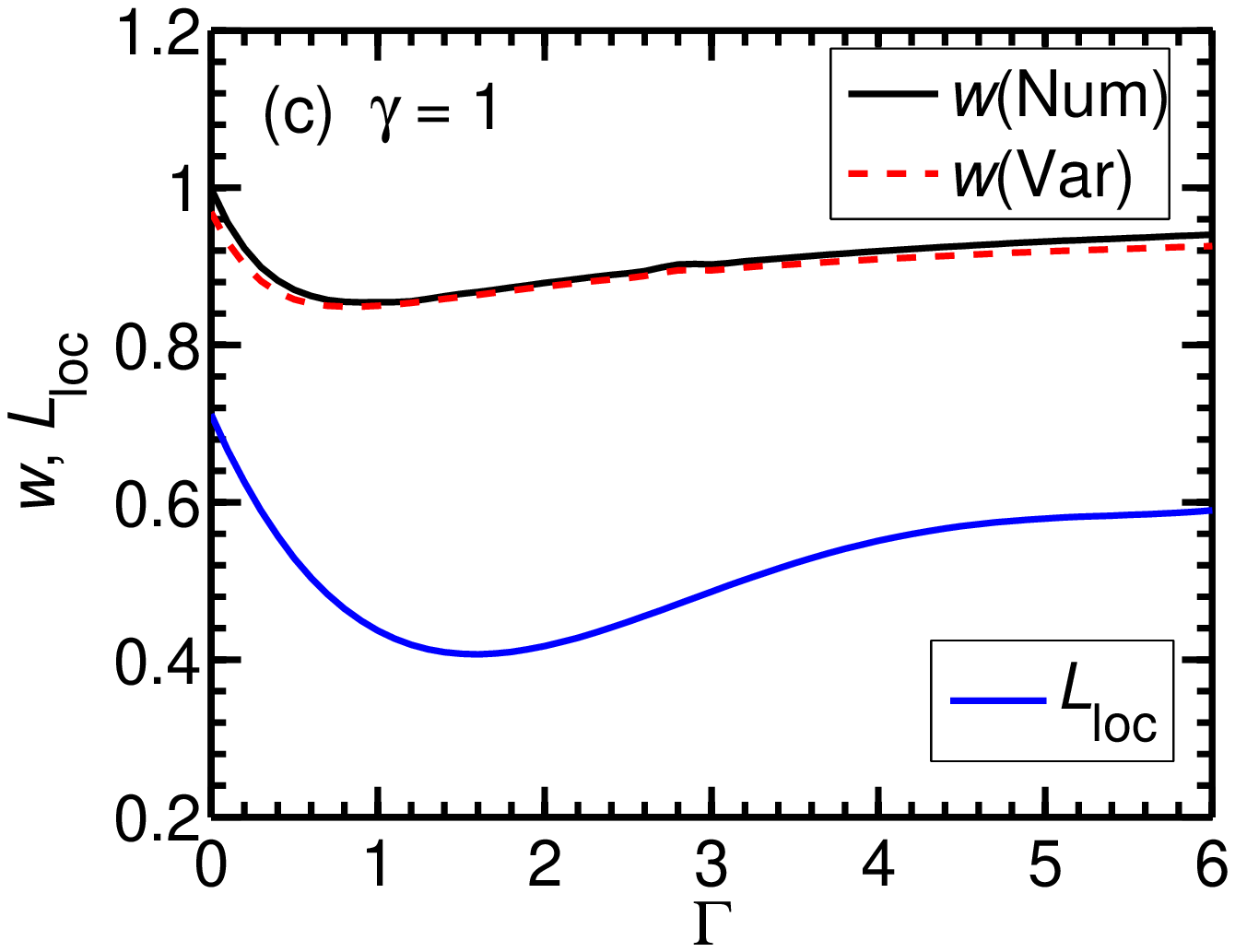}
\includegraphics[width=.49\linewidth,clip]{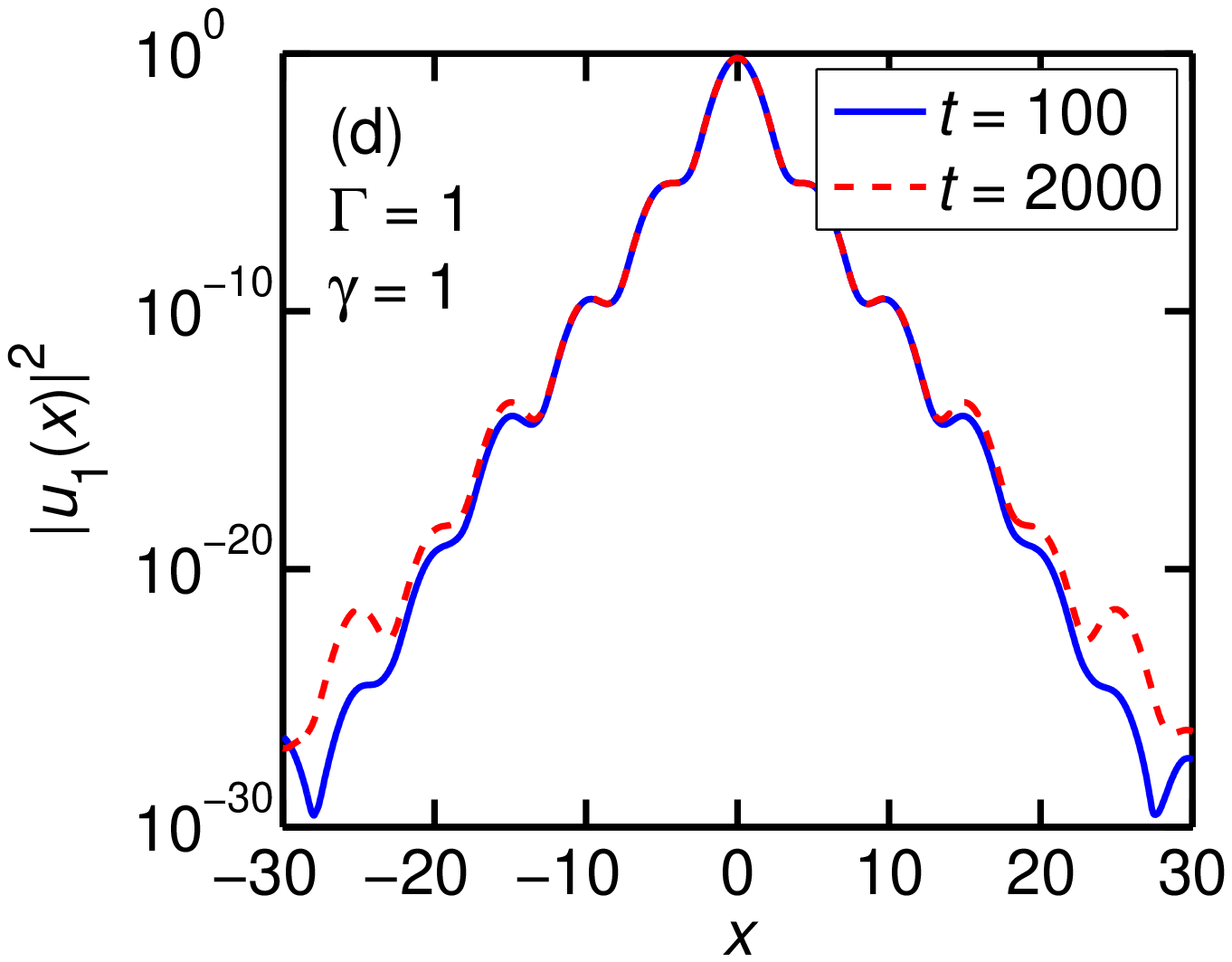}
\end{center}

\caption{(Color online) Numerical results for $\phi=\pi$(out of phase), $g=g_{12}=0, N_1=N_2=1$   (a) of the wave function $u_1$ and
density for $\gamma=\Gamma=4$, (b) of a $\Gamma-\gamma$ phase plot
showing Gaussian and non-Gaussian localization regions,
(c) of width and localization length for $\gamma=1$ and different $\Gamma$, and (d) density at large times $t$ for $\gamma=\Gamma=1$.        In (c) the variational results of widths are also shown.
 } \label{fig3}
\end{figure}

In the case of $\phi=\pi$ (out of phase),  from Eq. (\ref{WID1})
we find that a positive $\Gamma$
favors localization and   the set of Eqs. (\ref{WID1}) and (\ref{BET1}) has a real solution for width $w$ with arbitrary positive $\Gamma$ and $\gamma$.  In this case  the density profile of the stable localized state could be modulated with a  non-Gaussion shape { even within the single OL site}, as  illustrated in Fig. \ref{fig3} (a). From a numerical solution of  Eq.  (\ref{CGP1}),  we obtain the  phase diagram of Fig. \ref{fig3} (b) of $\gamma$ and $\Gamma$ showing the regions where the density profile is Gaussion and non-Gaussion.   To confirm this, a plot of the numerical  and variational   widths versus $\Gamma$ is presented in Fig. \ref{fig3} (c) for $\gamma= 1$ which shows  that a positive $\Gamma$ contributes a localization  of the  BEC because the width for $\Gamma>0$ is less than that for $\Gamma=0$. { So far we considered the central density of the localized states. A careful examination of the densities of the localized states reveals that
at large distances from the central region 
the  localized states always have a long exponential tail within both Gaussion and   non-Gaussion regimes.  In fact this is the most important earmark of Anderson localization. }
A measure of this tail can be given by a localization length  $L_{\mathrm{loc}}$ obtained by  fitting the density tail to the exponential function  $\sim\exp(-|x|/L_{loc})$ \cite{review-AL,localization-length}. The localization effect  of a positive $\Gamma$, however, will have a major influence on the exponential tail. For $\gamma= 1$, the effect of $\Gamma$ on the localization length $L_{loc}$ is presented in Fig. \ref{fig3} (c).   As expected, the localization effect of a nonzero $\Gamma$ makes  the localization length to be always smaller than  that for $\Gamma=0$. Furthermore, at large time scales, Fig. \ref{fig3} (d) illustrates that a subdiffusion occurs below a certain critical $\Gamma\times\gamma$, as analyzed numerically in Ref. \cite{subdiffusive} where  a subdiffusion appears  above a certain strength of nonlinearity.   Nevertheless, spreading induced by the subdiffusion is rather slow \cite{subdiffusive}. So, the localization lengths shown in Fig. \ref{fig3} (c) are meaningful.

Thirdly, we consider the stability of the solutions (\ref{SOL1})
and (\ref{SOL2}) by performing a standard linear stability analysis.
Introducing small fluctuations around the stationary solution ($R_0,
\phi_0$), $R'(t)=R(t)-R_0,\phi'(t)=\phi(t)-\phi_0$, and linearizing
Eqs. (\ref{PHI}) and (\ref{Nj}), a set of two
linear equations are obtained:
\begin{eqnarray}
\frac{d R'(t)}{dt}&=&F_R(R_0, \phi_0)R'(t)+F_\phi(R_0, \phi_0)\phi'(t),\\
\frac{d \phi'(t)}{dt}&=&G_R(R_0, \phi_0)R'(t)+G_\phi(R_0, \phi_0)\phi'(t),
\end{eqnarray}
where the  subscripts $R$ and $\phi$ denote a derivative with respect to the respective
variable. Assuming the solution of $R'(t)$ and $\phi'(t)$ in exponential form, $\sim \exp(\zeta t)$, the eigenvalue $\zeta$ is
given by 
\begin{align}
2\zeta&=F_R(R_0, \phi_0)+G_\phi(R_0, \phi_0) 
\pm\Bigl\{\big[F_R(R_0, \phi_0)  \nonumber \\
&-G_\phi(R_0, \phi_0)\big]^2
+4F_\phi(R_0, \phi_0)G_R(R_0, \phi_0)\Bigr\}^{1/2}.
\end{align}
From Eqs.   (\ref{PHI}) and (\ref{Nj}), we find $F_R(R_0, \phi_0)=0, G_\phi(R_0, \phi_0)=0,$ and
\begin{align}
& F_\phi(R_0, \phi_0)=\mp 2\Gamma \exp{\left(-\beta^2w^2\right)}, \label{Fphi}\\
& G_R(R_0, \phi_0)=N\left[\frac{g_{12}-g}{\sqrt{2\pi}w}\pm\frac{2\Gamma}{N}
\exp{\left(-\beta^2w^2\right)}\right],\label{GR}
\end{align}
which leads to the eigen-values
\begin{eqnarray}\label{eigen}
\zeta=\pm\left[F_\phi(R_0, \phi_0)G_R(R_0, \phi_0)\right]^{1/2}.
\end{eqnarray}
If the eigen-value $\zeta$ is purely imaginary, the stationary solutions denoted by Eqs. (\ref{SOL1}) and (\ref{SOL2}) are stable with respect to small perturbations. The constraints for stability are
\begin{eqnarray}
&& \frac{g_{12}-g}{\sqrt{2\pi}w}+\frac{2\Gamma}{N}
\exp{\left(-\beta^2w^2\right)}>0, \quad (\phi=0) \label{CON1}\\
&& \frac{g_{12}-g}{\sqrt{2\pi}w}-\frac{2\Gamma}{N}
\exp{\left(-\beta^2w^2\right)}<0. \quad (\phi=\pi) \label{CON2}
\end{eqnarray}
A straightforward conclusion from Eqs. (\ref{CON1}) and (\ref{CON2}) is that any stationary state is stable for  $g=g_{12}$.
In addition, the conditions for stability  are different for the in-phase and out-of-phase localized states when $g\ne g_{12}$.

\begin{figure}
\begin{center}
\includegraphics[width=.49\linewidth]{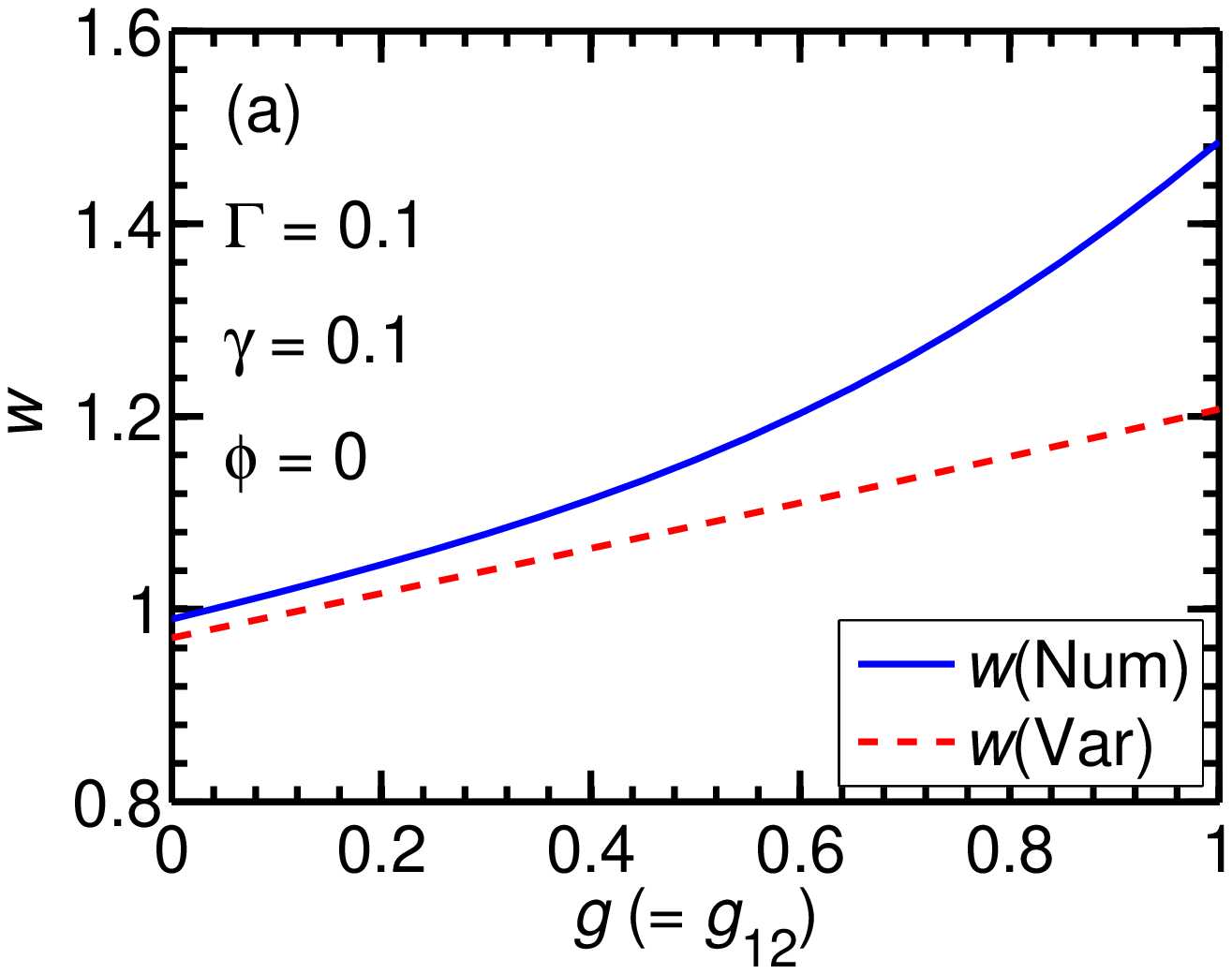}
\includegraphics[width=.49\linewidth]{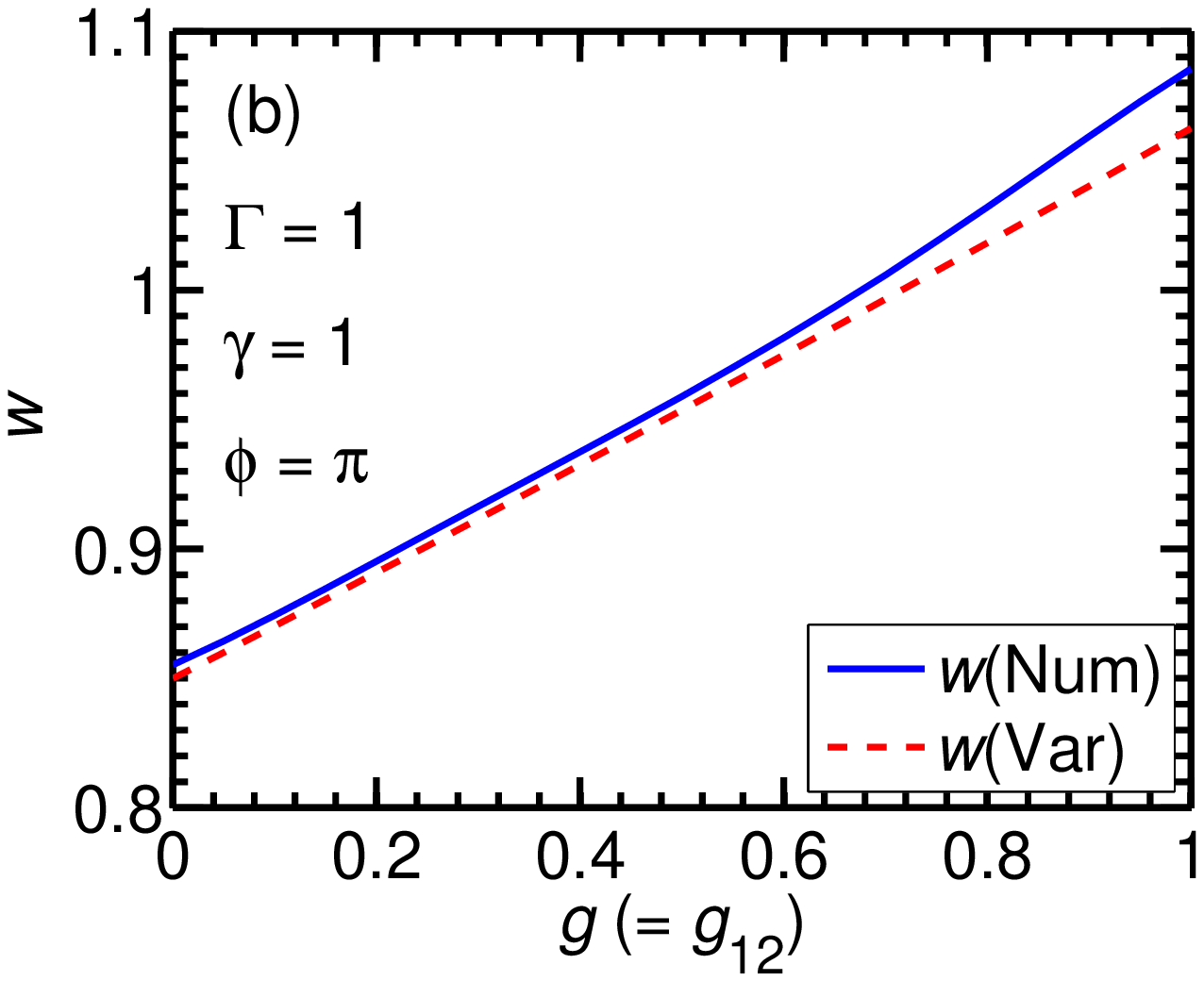}
\includegraphics[width=.49\linewidth]{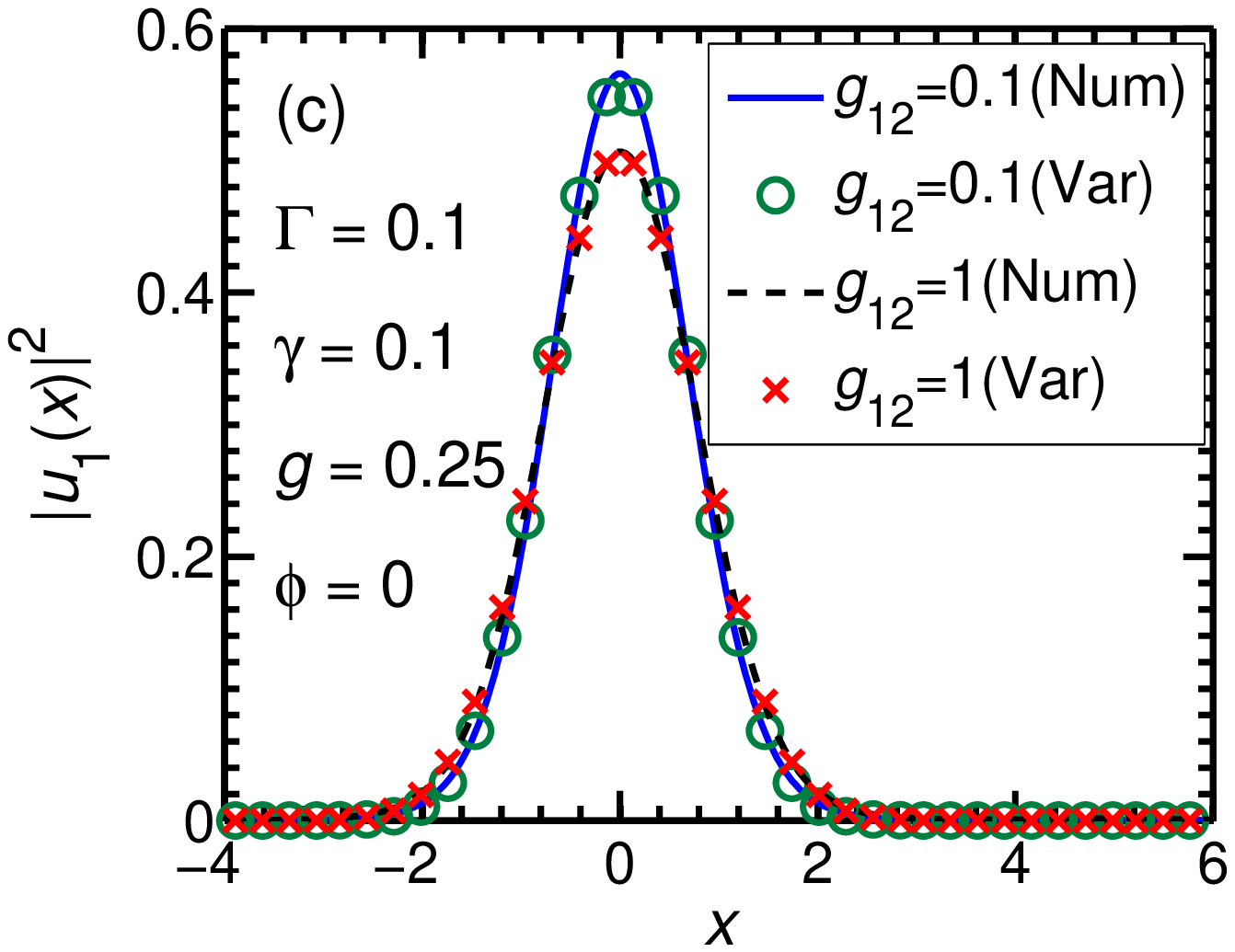}
\includegraphics[width=.49\linewidth]{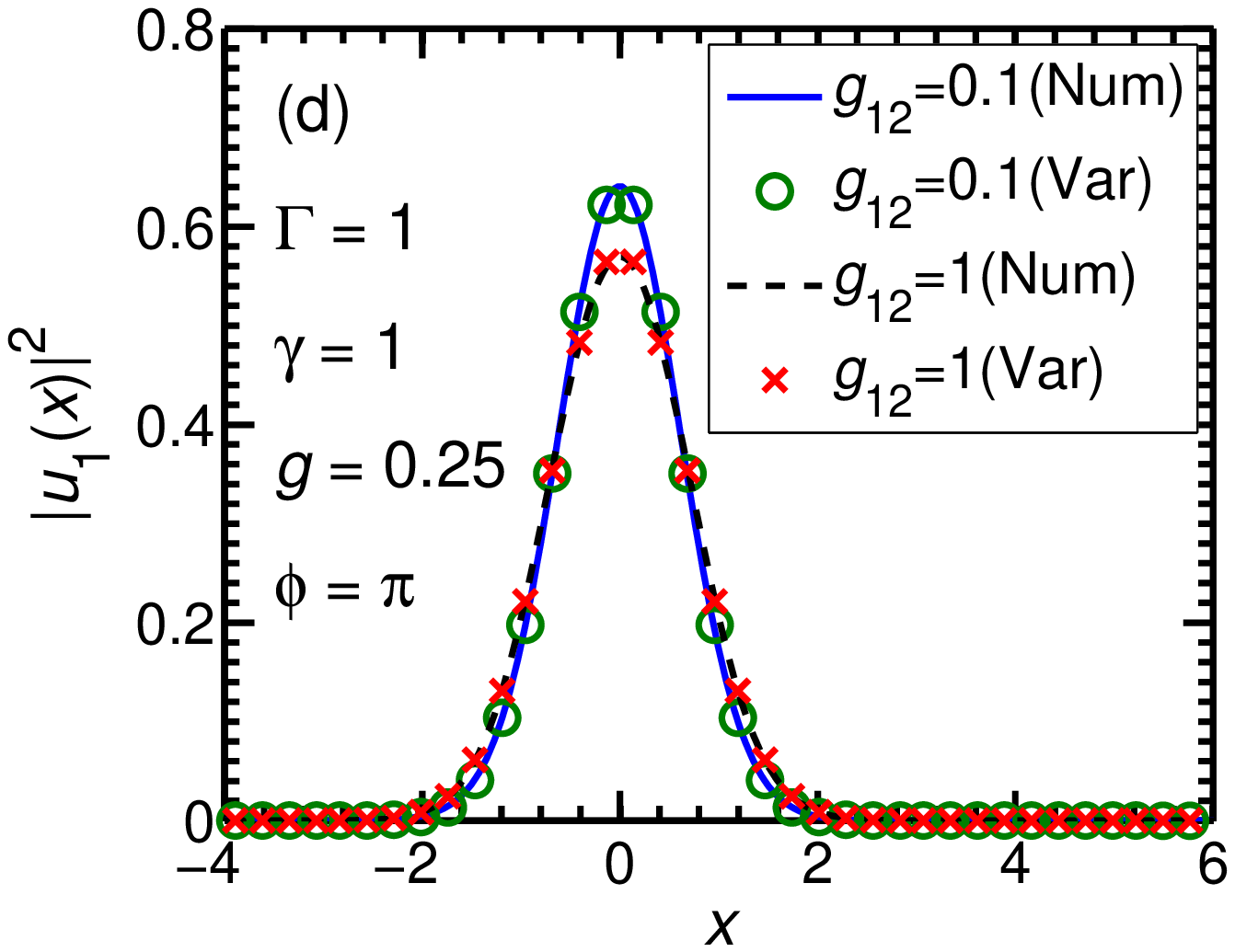}
\end{center}

\caption{(Color online) Numerical (Num)  and variational (Var) widths versus $g$ $(=g_{12})$  for (a) $\Gamma=\gamma= 0.1$ and $\phi=0$, (b) $\Gamma=\gamma= 1$ and $\phi=\pi$. Numerical (Num) and variational (Var) wave functions $|u_1|^2$ versus $x$ for  $g\neq g_{12}$ and for (c) $\Gamma=\gamma= 0.1, \phi=0$, and for (d) $\Gamma=\gamma= 1, \phi=\pi$.
 } \label{fig4}
\end{figure}

Now, let us investigate the effect of the positive nonlinearity on the localized states in presence of the SO and Rabi couplings. As shown by Eq. (\ref{WID1}), the stable localized states can exist within a range of parameter of $g,\,g_{12},\,\Gamma$ and $\gamma$ and this  has also been also confirmed by the numerical integration of Eq. (\ref{CGP1}).
The numerical  and variational  widths versus $g$ ($=g_{12}$) are plotted in Fig. \ref{fig4} (a) for $\phi=0, \gamma=\Gamma=0.1$ and (b) for $\phi=\pi, \gamma=\Gamma=1$. With these parameters, the localized states are confined in a single OL site. Figures \ref{fig4} (a) and (b) indicate that the numerical and variational widths increase monotonically as the nonlinearity (increase of repulsion) increases. The numerical results are slightly larger than the variational results because of the exponential tail of the localized state. If the nonlinearity is large enough, however, the localized states develop undulating tails occupying more than one OL site and cannot be described well by the Gaussian ansatz (\ref{ans}). For $g\neq g_{12}$, the typical numerical  and variational  densities $|u_1|^2$ versus $x$ are illustrated in Fig. \ref{fig4}  
 (c) for $\phi=0$ and (d) for $\phi=\pi$. The parameters are chosen to meet the stability criteria   (\ref{CON1}) and (\ref{CON2}). The stability for those localized states are tested by suddenly changing OL's intensity $s_1$ from 10 to 9.5 and continually running the real-time program. The localized states are again found to be stable against the small perturbation.

\section{DYNAMICS OF LOCALIZED STATE} \label{IIII}

\begin{figure}
\begin{center}
\includegraphics[width=\linewidth]{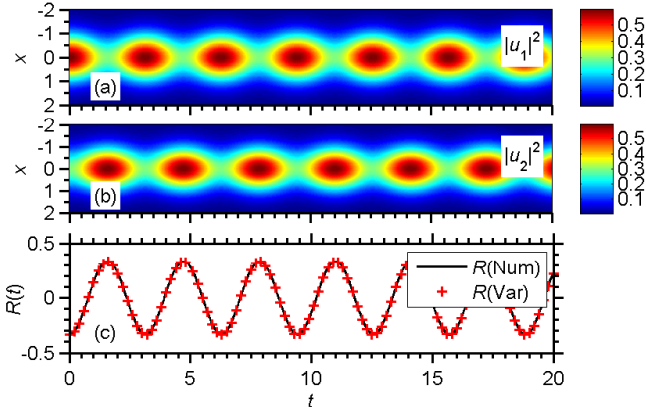}
\end{center}

\caption{(Color online) Numerical density profile versus time $t$ (a) for $|u_1|^2$ and (b) for $|u_2|^2$. (c) Numerical (Num) and variational (Var) atom transfer ratio $R(t)$ versus time $t$.  The parameters are $\Gamma=1$, $\phi(0)=0$, $g=g_{12}=\gamma=0$, and $ N_2(0)=0.5, N_1(0)=1, R(0)=[N_2(0)-N_1(0)]/[N_2(0)+N_1(0)]=-0.3333$.
 } \label{fig5}
\end{figure}

For the stationary states studied so far one must have $N_1=N_2$. A  little imbalance between $N_1$ and $N_2$  leads to periodic atom transfer between two components. The localized states may exist in that case although the wave functions change with time with periodic atom transfer between components.

In order to get a further insight into the effects of the the coefficient $\gamma$ and $\Gamma$  on the localized states, we now study some dynamics of the noninteracting and weakly interacting BEC. As shown by Eq. (\ref{Nj}), Rabi-coupling strength $\Gamma$ plays an important role for the atom transfer between components. We investigate the effects of SO coupling 
$\gamma$ and Rabi coupling $\Gamma$  on the atom transfer ration  $R(t)$. If $\gamma=0$ and $g=g_{12}$, Eqs.   (\ref{PHI}) and  (\ref{Nj}) become
  \begin{eqnarray}\label{Nj1}
\dot{R}(t)&=&-2\Gamma\sin\phi(t)\sqrt{1-R^2(t)},\\\label{PHI1}
\dot{\phi}(t)&=&\frac{2\Gamma R(t)\cos\phi(t)}{\sqrt{1-R^2(t)}},
\end{eqnarray}
with solutions
\begin{eqnarray}\label{rt}
R(t)&=&A\cos(2\Gamma t+B),\\\label{phit}
\left|\cos\phi(t)\right|&=&\frac{\sqrt{1-A^2}}{\sqrt{1-R^2(t)}},\quad (R<1),
\end{eqnarray}      
where $A$ and $B$ are integration constants, which are determined by the initial values   $R(0)$ and $\phi(0)$. Equation (\ref{rt}) shows that the period of atom transfer is determined solely by  $\Gamma$.  From Eqs. (\ref{rt}) and (\ref{phit}), we can deduce that the integration constants $A$ and $B$ change periodically with $\phi(0)$. The minimum of $A$ is $A=R(0)$ corresponding to $\phi(0)=0$, and the maximum is $A=1$ corresponding to $\phi(0)=\pi/2$. Notice that, theoretically, equations (\ref{PHI1}) or (\ref{phit}) has the { singular points $R(t)=\pm 1$}, that should be encountered if $\phi(0)=\pi/2$. {  However, the numerical integration of Eq. (\ref{CGP1}) reveals that $R(t)=\pm 1$ can be achieved when $\phi(0)=\pi/2$, and the atom transfer can go on in this  case with $R(t)$ oscillating periodically between $\pm 1$.}

Contour plots of  numerical density profiles versus time $t$ are displayed in Fig. \ref{fig5} (a) for $|u_1|^2$, Fig. \ref{fig5} (b) for $|u_2|^2$ for $g=g_{12}=0$. The periodic atom transfer between
the components is clear in these plots. A quantitative  measure of this oscillation is given by the plot of numerical and variational estimates of atom transfer ratio 
 $R(t)$ versus $t$ in Fig. \ref{fig5} (c).
 The variational $R(t)$ is obtained by a numerical integration of Eqs. (\ref{Nj1}) and (\ref{PHI1}) with the fourth order Runge-Kutta method. To obtain the numerical $R(t)$, we first obtain  the stationary states employing the imaginary-time method solving the Eqs. (\ref{CGP1}) and (\ref{pot}) with $\gamma=\Gamma=0$
and $N_j(0)$ such that  $R(0)=[N_2(0)-N_1(0)]/[N_2(0)+N_1(0)]$ and $\phi(0)=\phi_{20}(0)-\phi_{10}(0)=0$.   Successively, at $t = 0$, we employ the real-time propagation of  Eq. (\ref{CGP1})  with the same parameters   and just changing $\Gamma$ from 0 to 1.  The atom transfer between components start for a nonzero $\Gamma$ and a numerical $R(t)$  is obtained by calculating $R(t)=[N_2(t)-N_1(t)]/[N_2(t)+N_1(t)]$ with $N_j(t)=\int_{-\infty}^{-\infty}|u_j(x,t)|^2dx$.

\begin{figure}
\begin{center}
\includegraphics[width=\linewidth]{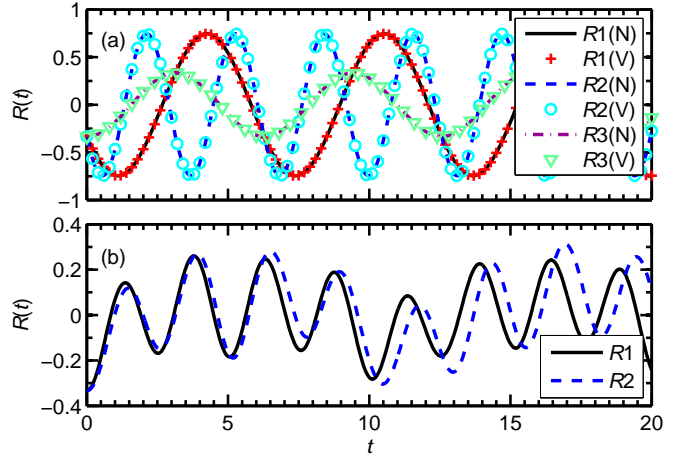}
\end{center}

\caption{(Color online) (a) Numerical (N) and variational (V) atom transfer ratio $R(t)$ versus time $t$ for $\gamma=0$ in different cases:  $R1$ for $g=g_{12}=0,\; \phi(0)=\pi/4,\;\Gamma=0.5$;  $R2$ for $g=g_{12}=-1,\; \phi(0)=\pi/4,\;\Gamma=1$; and $R3$ for $g=g_{12}=-0.5,\; \phi(0)=0,\;\Gamma=0.5$. (b) Numerical atom transfer ratio $R(t)$ versus time $t$ with $\phi(0)=0$ in different cases:  $R1$ for $g=g_{12}=0,\; \Gamma=\gamma=1$;  $R2$ for $g=0,\;g_{12}=-1,\; \Gamma=\gamma=1$. In all cases $ N_2(0)=0.5, N_1(0)=1, R(0)=[N_2(0)-N_1(0)]/[N_2(0)+N_1(0)]=-0.3333$.
 } \label{fig6}
\end{figure}

Next we study the atom transfer between components for $g=g_{12}\neq0$. In Fig. \ref{fig6} (a) we plot numerical and variational results for the atom transfer ratio $R(t)$ in several cases  
{ for $\gamma=0$}. The numerical results are in  good  agreement with variational Eqs. (\ref{Nj1}) and (\ref{PHI1}) which shows that the period of $R(t)$ is related to only $\Gamma$, and the effect of $\phi(0)$ on $A$ and $B$ is larger than the effect of $R(0)$. Further investigations show that, if $\gamma\neq0$, the density profiles may be non-Gaussion, and the variational equations (\ref{Nj})--(\ref{BET}) are no longer valid  although the atom transfer between two components can take place as demonstrated by lines  $R1$ and $R2$ in Fig. \ref{fig6} (b).
These plots   in Fig. \ref{fig6} (b) indicate that the amplitude of $R$ changes periodically.

\section{SUMMARY}
\label{IIIII}

Using the numerical solution and variational approximation of the time-dependent coupled mean-field GP equations with two pseudo spin components, we studied the localization of the noninteracting and weakly interacting Bose-Einstein condensates with SO and Rabi couplings loaded in the quasiperiodic bichromatic OL potential (\ref{pot}). We use the set of binary GP equations (\ref{CGP1})
that predicts accurately the evolution of the atom transfer ratio $R(t)$, phase difference $\phi(t)$, and width $w$. The  variational results leading to many physical insights
are compared  with the  numerical results of the mean-field model.
Stationary localized states of the model correspond to the same number of atoms ($N_1=N_2$) in two components. Nonstationary localized states with periodic atom transfer between components can be achieved for different number of atoms ($N_1\ne N_2$).
 In the case of $\gamma\times\Gamma=0$, the  density profiles of the two stationary localized states   are symmetrical, and are not related to the phase difference $\phi$, SO coupling $\gamma$ and Rabi coupling $\Gamma$. In the case of $\gamma\times\Gamma\neq0$, the width of the stationary state should depend on the the phase difference and the SO and Rabi couplings because of the interaction between $\gamma$ and $\Gamma$. It is found that the interaction between the SO coupling and Rabi coupling may favor a localization or delocalization
  depending on the the phase difference between the two localized states. If $g = g_{12}$, a linear stability analysis shows that any stationary state is stable.  We find that the BEC localized states always have a long exponential tail. In the case of $\phi=\pi$, the localization effect of a positive $\Gamma$ has a major influence on the localization length. We also studied some dynamics of the localized states with the atomic population imbalance, and find $\Gamma$ and the initial phase difference play an important role for the atom transfer. Either in view of understanding the dynamic evolution or in view of the practical application, these properties are important. We hope that the present work will motivate new studies, specially experimental ones on the localization of BEC with the SO coupling.

\acknowledgments

FAPESP and CNPq (Brazil) provided partial support. Y.  Cheng
undertook this work is supported by National Natural Science Foundation of China Grant No. 11274104 and Provincial Natural Science Foundation of Hubei Grant No. 2011CDA021.

\hskip .2 cm

\newpage



\begin{thebibliography}{10}

\bibitem{nature-471-83} Y.-J. Lin, K. Jim\'{e}nez-Garc\'{\i}a, and I. B. Spielman, Nature (London) {\bf 471}, 83 (2011).

\bibitem{nature-494-49} V. Galitski and Ian B. Spielman,
 Nature   (London) {\bf 494}, 49 (2013).

\bibitem{experimental-scheme} J. Higbie and D. M. Stamper-Kurn,  Phys. Rev. Lett. {\bf 88}, 090401 (2002); T. L. Ho and S. Zhang,  Phys. Rev. Lett. {\bf 107}, 150403 (2011); Y. Deng, J. Cheng,  H. Jing, C. P. Sun, and S. Yi, Phys. Rev. Lett. {\bf 108}, 125301 (2012); J. Radic, T. A. Sedrakyan, I. B. Spielman, and V. Galitski, Phys. Rev. A {\bf 84}, 063604 (2011).

\bibitem{Fermi-gas} P. Wang, Z.-Q. Yu, Z. Fu,  J. Miao, L. Huang, S. Chai, H. Zhai, and J. Zhang,  Phys. Rev. Lett. {\bf 109}, 095301 (2012); L. W. Cheuk, A. T. Sommer, Z. Hadzibabic, T. Yefsah, W. S. Bakr, and M. W. Zwierlein,  Phys. Rev. Lett. {\bf 109}, 095302 (2012).

\bibitem{Bose-gas} J. Y. Zhang, S.C. Ji, and Z. Chen {\it et al.}, Phys. Rev. Lett. {\bf 109}, 115301 (2012); M. Aidelsburger, M. Atala, and S. Nascimb\'{e}ne {\it et al.}, Phys. Rev. Lett. {\bf 107}, 255301 (2011); Z. Fu, P. Wang, and S. Chai,  L. Huang, and J. Zhang, Phys. Rev. A {\bf 84}, 043609 (2011); C. Qu, C. Hamner,  M. Gong, C. Zhang, and P. Engels, Phys. Rev. A {\bf 88}, 021604 (2013).

\bibitem{superfluidity} D. W. Zhang, J. P. Chen,  C. J. Shan, Z. D. Wang, and S. L. Zhu, Phys. Rev. A {\bf 88}, 013612 (2013); Q. Zhu, C. Zhang and B. Wu, Europhys. Lett. {\bf 100}  50003 (2012).

\bibitem{vortex} X. F. Zhou, J. Zhou, and C. Wu, Phys. Rev. A {\bf 84}, 063624 (2011).

\bibitem{soliton} 
 Y. Xu, Y. Zhang, and B. Wu, Phys. Rev. A {\bf 87}, 013614 (2013); O. Fialko, J. Brand, and U. Z{\"u}licke, Phys. Rev. A {\bf 85}, 051605(R) (2012).

\bibitem{BCS-BEC} M. Gong, G. Chen,  S. Jia, and C. Zhang, Phys. Rev. Lett. {\bf 109}, 105302 (2012); M. Gong, S. Tewari, and C. Zhang, Phys. Rev. Lett. {\bf 107}, 195303 (2011); H. Hu, L. Jiang,  X. J. Liu, and H. Pu, Phys. Rev. Lett. {\bf 107}, 195304 (2011); L. Han and C. A. R. S{\'a} de Melo, Phys. Rev. A {\bf 85}, 011606 (2012).

\bibitem{GP-equation1} Y. Zhang, L. Mao, and C. Zhang,  Phys. Rev. Lett. {\bf 108}, 035302 (2012).

\bibitem{GP-equation2} T. D. Stanescu, B. Anderson, and V. Galitski, Phys. Rev. A {\bf 78}, 023616 (2008); J. Larson and E. Sj{\"o}qvist, Phys. Rev. A {\bf 79}, 043627 (2009).

\bibitem{localized-modes} L. Salasnich and B. A. Malomed, Phys. Rev. A {\bf 87}, 063625 (2013). 

\bibitem{use-GPE}Y. Li, G. I. Martone, L. P. Pitaevskii, and S. Stringari, Phys. Rev. Lett. {\bf 110}, 235302 (2013);
Y. Zhang and C. Zhang, Phys. Rev. A {\bf 87}, 023611 (2013).


\bibitem{use-GPE2} C. Wang, C. Gao,  C. M. Jian, and H Zhai, Phys. Rev. Lett. {\bf 105}, 160403 (2010);   T. Kawakami, T. Mizushima, and K. Machida, Phys. Rev. A {\bf 84}, 011607(R) (2011); A. Aftalion and P. Mason, Phys. Rev. A {\bf 88}, 023610 (2013).



\bibitem{anderson} P. W. Anderson, Phys. Rev. {\bf 109}, 1492 (1958).

\bibitem{billy} J. Billy, V. Josse, and Z. Zuo {\it et al.}, Nature
 (London) {\bf 453}, 891(2008).

\bibitem{roati} G. Roati, C. D'Errico, and L. Fallani {\it et al.}, Nature (London) {\bf 453}, 895 (2008).

\bibitem{PRL-98-130404} L. Fallani, J. E. Lye, V. Guarrera,  C. Fort, and M. Inguscio, Phys. Rev. Lett. {\bf 98}, 130404 (2007).

\bibitem{AL-3D1}S. S. Kondov, W. R. McGehee, J. J. Zirbel, and B. DeMarco, Science {\bf 334}, 66 (2011).

\bibitem{AL-3D2} F. Jendrzejewski, A. Bernard, and K. M\"uller {\it et al.}, Nature Phys. {\bf 8}, 398 (2012).


 

\bibitem{PRA-83-023620}{
B. Damski, J. Zakrzewski, L. Santos, P. Zoller, and M. Lewenstein,  \prl {\bf 91}, 080403 (2003);
L. Sanchez-Palencia, D. Cl\'ement, P. Lugan, P. Bouyer, G. V. Shlyapnikov, and A. Aspect, \prl 
{\bf 98}, 210401 (2007);
S. E. Skipetrov, A. Minguzzi, B. A. van Tiggelen, and B. Shapiro, 
\prl {\bf 100}, 165301 (2008);
A. Yedjour and  B. A. Van Tiggelen, 
 Eur Phys J D {\bf 59}, 249 (2010);
M. Piraud, L.  Pezze, and  L.  Sanchez-Palencia, 
 Europhys Lett {\bf 99}, 50003 (2012); 
  New J. Phys. {\bf 15}, 075007 (2013);} J. Biddle, B. Wang, D. J. 
Priour, and S. DasSarma, Phys. Rev. A {\bf 80}, 021603 (2009); M. 
Larcher, F. Dalfovo, and M. Modugno, Phys. Rev. A {\bf 80}, 053606 
(2009); M. Modugno, New J. Phys., {\bf 11}, 033023 (2009); Y. Cheng and 
S. K. Adhikari, Phys. Rev. A {\bf 83}, 023620 (2011); Phys. Rev. A {\bf 
84}, 023632 (2011); Phys. Rev. A {\bf 84} 053634 (2011); Phys. Rev. A 
{\bf 82}, 013631 (2010);
 S. K. Adhikari and L. Salasnich, \pra {\bf 80},
 023606 (2009); M Takahashi, H Katsura, M Kohmoto, and T Koma,
New J. Phys. {\bf 14}, 113012 (2012). 



\bibitem{AL-in-SOC1} L. Zhou, H. Pu, and W. Zhang, Phys. Rev. A {\bf 87}, 023625 (2013).

\bibitem{AL-in-SOC2} M. J. Edmonds, J. Otterbach, R. G. Unanyan {\it et al.}, New J. Phys. {\bf 14}, 073056 (2012).

\bibitem{variational}  V. M. P\'erez-Garc\'ia, H. Michinel,  J. I. Cirac, M. Lewenstein, and P. Zoller, Phys. Rev. A {\bf 56}, 1424 (1997);   Y.  Cheng, R. Z. Gong  and H. Li, Opt. Express {\bf 14,} 3594(2006); B. A. Malomed, Prog. in Optics {\bf 43}, 69 (2002).

\bibitem{review-AL} P. Bouyer, Rep. Prog. Phys. {\bf 73}, 062401 (2010);  L. Sanchez-Palencia and M. Lewenstein, Nature Phys. {\bf 6}, 87 (2010); L. Fallani, C. Fort and M. Inguscio, Adv. At. Mol.  Opt. Phys. {\bf 56}, 119 (2008).



 \bibitem{7}Y. A. Bychkov and E. I.  Rashba,  J. Phys. C {\bf 17}, 6039 (1984).
\bibitem{8} G. Dresselhaus, Phys. Rev. {\bf 100}, 580 (1955). 


\bibitem{1dsala}L. Salasnich, A. Parola, and L. Reatto, \pra
{\bf 65},   043614   (2002);
C. A. G. Buitrago and  S. K. Adhikari, J. Phys. B {\bf 42}, 215306 (2009).


\bibitem{opt}S. Blatt {\it et al.}, \prl {\bf 107}, 073202 (2011).




\bibitem{Feshbach} S. L. Cornish, N. R. Claussen,   J. L. Roberts, E. A. Cornell, and C. E. Wieman  Phys. Rev. Lett. {\bf 85}, 1795 (2000); M. Theis, G. Thalhammer, and K. Winkler {\it et al.}, Phys. Rev. Lett. {\bf 93}, 123001 (2004); S. E. Pollack, D. Dries, and M. Junker {\it et al.},  Phys. Rev. Lett. {\bf 102}, 090402 (2009); S. Inouye, M. R. Andrews, and J. Stenger {\it et al.}, Nature (London) {\bf 392}, 151 (1998).

 





\bibitem{localization-length} N. F. Mott,  J. nonCryst. Solids {\bf 1}, 1 (1968).

\bibitem{subdiffusive}S. Flach, D. O. Krimer, and Ch. Skokos, Phys. Rev. Lett. {\bf 102}, 024101 (2009); A. S. Pikovsky and D. L. Shepelyansky, \prl {\bf 100}, 094101 (2008).

\end{thebibliography}
\end{document}